\documentclass[twocolumn,groupedaddress,showpacs,floatfix]{revtex4}
\usepackage{color}
\usepackage{graphicx}
\usepackage{subfigure}
\newcommand{\sLT}{\mbox{$\sigma_{LT}$}}

\newcommand{\ndelta}{\ifmmode {N\!\!\rightarrow\!\!\Delta} \else{$N\!\!\rightarrow\!\!\Delta$}\fi}

\newcommand{\be}{\begin{equation}}
\newcommand{\ee}{\end{equation}}
\newcommand{\beq}{\begin{eqnarray}}
\newcommand{\eeq}{\end{eqnarray}}
\newcommand{\bw}{\begin{widetext}}
\newcommand{\ew}{\end{widetext}}

\newcommand{\gevc}{\mbox{GeV$^2$}}
\newcommand{\GNdelta}{\mbox{$\gamma^* N \to \Delta$}}

\begin{document}

\title{The Shape of Hadrons}

\author{Constantia Alexandrou}
\affiliation{Department of Physics, University of Cyprus, P.O. Box 20537, 1678
Nicosia, Cyprus and The Cyprus Institute, 15 Kypranoros St., Nicosia, Cyprus}

\author{Costas~N.~Papanicolas \footnote{corresponding author, email address:
cnp@cyi.ac.cy}}
\affiliation{The Cyprus Institute, 15 Kypranoros St., Nicosia, Cyprus and
Department of Physics, University of Athens, Athens, Greece}

\author{Marc Vanderhaeghen}
\affiliation{Institut f\"ur Kernphysik, Johannes Gutenberg-Universit\"at,
D-55099 Mainz, Germany}

\date{\today}

\begin{abstract}
 This colloquium addresses the issue of the shape
of hadrons and in particular that of the proton. The concept of
shape in the microcosm is critically examined.
Special attention is devoted to properly define the meaning of shape
 for bound-state systems
of near massless quarks.
The  ideas that lead  to the
expectation of non-sphericity in the shape of hadrons,
the calculations that predict it,
and the experimental information obtained from recent high-precision
measurements are examined. Particular emphasis is given to the
study of the electromagnetic transition between the nucleon and its
first excited state, the $\Delta$(1232)-resonance. The experimental evidence is
is critically examined and compared with  lattice calculations,
as well as with effective-field
theories and phenomenological models.
\end{abstract}

\maketitle
\tableofcontents

\section{Introduction}
\label{sec:intro}

Hadrons are the smallest material objects in the universe known to
be of finite size.  The building blocks of the Standard Model,
including leptons and quarks, are not known to have finite size with
the smallest scale set by the experimental limit on the size of the electron,
which is  smaller  than $10^{-18}$m in diameter~\cite{Gabrielse:2006gg}. Hadrons
are distinguished in two families: mesons, which are made out of a
quark and an antiquark pair and baryons, which are made out of three
quarks. Quantum Chromodynamics (QCD) does not exclude the
possibility of other forms of hadronic matter such as di-baryons or pentaquarks,
but none of these have been found thus far. The typical scale of a hadron
radius is
set by the well known charge radius of the proton, which is equal
to $ 0.8768 \, (69) \, \times 10^{-15}$m~\cite{Mohr:2008fa}.
The very concept of size, both classically and quantum mechanically,
raises the issue of shape and it is therefore natural to inquire about the
shape of
hadrons. The shape of hadrons concerns microscopic objects
at the scale of a femtometer ($10^{-15}$m).

Inquiring about the shape of a subatomic particle is equivalent to
raising the question whether the distribution of its constituents or
some of the extensive and therefore distributed properties, such as
mass or charge, deviate from a spherical distribution, which is
assumed to be the default distribution. However, hadrons are objects
which are understood only within a quantum mechanical,
relativistic framework. It is thus  necessary to reexamine
critically the concept of size and shape in the context of quantum
mechanics and relativity before we  address the question of the
shape of hadrons. In parallel, we will also address the issue
of how sizes and shapes are determined for particles of the
microcosm.

Knowledge of the shape of the fundamental building blocks of the
Universe,  is not a curiosity, although it certainly comes close to
being an example of  the Aristotelian claim of the intrinsic human
need to ``know".  Experience from the determination and subsequent
understanding of shapes of other objects
 in the microcosm such as those of atoms and
nuclei, shows that this line of investigation is particularly
fertile for the understanding of the interactions of their
constituents amongst themselves and the surrounding medium. 
For hadrons this means the interquark interaction and
the quark - gluon dynamics.

While the theoretical foundations describing hadrons as the
smallest objects in the universe to which size and shape can be
attributed are solid, the empirical knowledge concerning shape is
limited and derives only from the detailed study of the transition
to the first excited state of the proton, the $\Delta^+(1232)$
resonance~\cite{Papanicolas:2007zz}. It is interesting to observe that while the determination
that hadrons have size emerged early on, through the seminal work of
R. Hofstadter and collaborators~\cite{hofs55}
and played a leading role in guiding hadron research ever since,
the determination of shape has been elusive and continues to be very limited.

In the rest of this section we will review the development of the
concepts of size and shape in classical and quantum mechanics with
and without relativity, so as to establish the appropriate language
needed to discuss the topic of ``Shape of Hadrons".
This is necessary as hadrons are systems 
requiring a relativistic quantum mechanical description.
In doing so, we will
provide some historical background on how these concepts have
developed.

\subsection{Historical Development}
\label{sec:history}

The issue of shape of subatomic particles arose most acutely  in the
case of nuclear physics. It is interesting to observe that
historically the issue concerning the shape of atoms being non
spherical never caused much surprise, perhaps because of the
planetary (Rutherford) model, which intrinsically invokes non-spherical shapes.
The discovery by Rabi and collaborators~\cite{Kellogg:1939zz}
that the deuteron had a static quadrupole moment and therefore its shape
 was not spherical
was regarded as a major surprise. The discovery of deformation in
the deuteron, and nuclei in general, was interpreted correctly as
arising due to the existence of non-central (tensor) forces among
nucleons. Shortly afterwards, Gerjuoy and Schwinger proposed that
trinucleon deformation (e.g. $^{3}$H) could resolve some peculiarities
 in the spectroscopy of those systems~\cite{Gerjuoy:1942zz}.
This conjecture proved
to be wrong - the effects were eventually understood to be due to
mesonic degrees of freedom. The deuteron and trinucleon cases
dramatically showed that understanding the shape of a subatomic particle
requires a detailed knowledge of its constituents and it  provides important
information for their  dynamics. Following the success of Rabi,
 the establishment and quantification of deformation through the measurement of the
electric quadrupole moment of a particle was widely employed
to map the systematics of deformation in atomic nuclei.

Unfortunately a number of misconceptions arose as a result of the
successful use of the determination of quadrupole moments in
inferring deviations from spherical shapes for atomic nuclei. The
fact that the measurement of a quadrupole moment is possible only
for systems (particles) possessing spin equal to one or bigger led
incorrectly to the belief that the shape of particles possessing
spin 0 or 1/2 cannot be determined. The  impossibility
to measure a quadrupole moment of such particles was mistakenly interpreted
as signifying a spherical shape. It took more than two decades
before this issue was clarified, primarily through the work of P.
Brix and coworkers~\cite{Brix:1977dx}. The realization that the determination
 of the intrinsic shape of a
system is quite distinct from the ability to measure its quadrupole moment
helped the field develop. Nuclear physicists developed new techniques
 to measure shapes that were
also applicable to nuclei of spin 0 or 1/2. In cases where the rigid rotor
(shape) approximation could be made, their excitation spectrum
could be used to reconstruct the density~\cite{Hersman:1986zz}.

\begin{figure}[h]
\vspace*{-9.5cm}
\hspace*{1cm}\includegraphics[width=1.4\linewidth,height=1.8\linewidth]{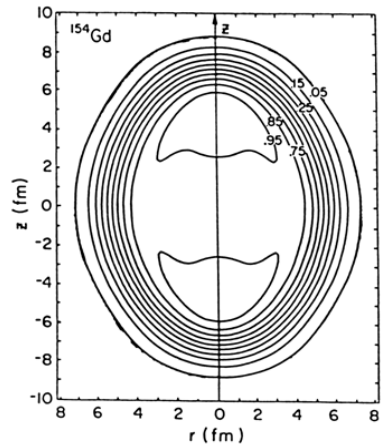}
\caption{Tomographic view of the deformed nucleus $^{154}$Gd derived
from the study of the rotational bands of this nucleus using
electron scattering~\cite{Hersman:1986zz}. The plot shows the contours
of equal charge density revealing its shape. This spin zero
nucleus has a vanishing quadrupole moment and therefore what is
revealed through this reconstruction is its intrinsic shape.}
\label{fig:Gd154_shape}
\end{figure}

\begin{figure}
\vspace*{-8.cm}
\hspace*{1cm}\includegraphics[width=1.7\linewidth,height=1.6\linewidth]{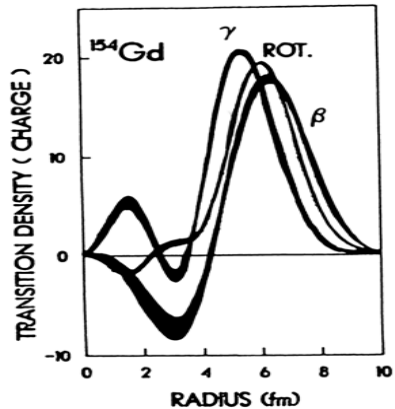}
\caption{The transition densities characterizing the moving charges
involved in excitations in  the deformed nucleus $^{154}$Gd, derived
from the study of this nucleus using electron scattering
~\cite{Hersman:1986zz}, demonstrate that vibrations along the two axes of
the ellipsoidal shape have different spatial extent.}
\label{fig:Gd154_trd}
\end{figure}

In the seventies and eighties high resolution electron scattering,
using this technique was used to map the shapes of the deformed
nuclei in the rare earth and actinide regions of the periodic table.
A superb example of the finesse of this tool is shown in
Figs.~\ref{fig:Gd154_shape} and \ref{fig:Gd154_trd}. The reconstructed
ground state charge density of the $^{154}$Gd  spin zero nucleus~\cite{Hersman:1986zz} is mapped
with high precision and the isodensity contours reveal vividly its
deformation, as is seen in Fig.~\ref{fig:Gd154_shape}.
Similarly the transition densities of $^{154}$Gd, shown in Fig.~\ref{fig:Gd154_trd}
provide a vivid geometric representation of the shape oscillations
of the system along the long and short axes of symmetry again
revealing its non-spherical charge distribution~\cite{Hersman:1986zz}.

The understanding of the shape of nuclei led to a better
understanding of nuclear structure and an appropriate language to
describe a number of important nuclear phenomena. The development of
the formalism of shape oscillations  viewed as normal
modes of the oscillating quantum liquid, nuclear matter, was
a crucial milestone in the field of nuclear physics and indeed of
physics. Modes of shape oscillations such as the ``giant dipole
oscillation'', the ``breathing mode'' or the ``scissor's mode"~\cite{Bohle:1983sx} have
yielded valuable knowledge on a number of parameters characterizing
nuclear matter. For instance, its compressibility, 
a parameter of critical importance in the understanding of supernova
explosions, is derived from the study of the breathing mode of
nuclei. Nuclear shapes and shape oscillations have also led to
paradigms, which are driving the development in other fields of
physics, such as the observation of a ``scissor's mode'' in Bose-Einstein condensates
in low temperature physics~\cite{stringari,marago}.

It may appear from the preceding introductory comments
that to inquire about the shape of an object possessing size is an obvious
undertaking. However, it took
more than twenty five years from the indication of the finite size
of the proton to the inquiry about its shape. The conjecture that
hadrons would have non-spherical amplitudes was first made by
Glashow in 1979 on the basis of non-central (tensor) interactions
between quarks~\cite{Glashow:1979gp}. Glashow argued that this would
resolve a number of inconsistencies that QCD was facing at the time
if the constrain of sphericity of the shape of hadrons was relaxed.
The conjecture  of non-spherical hadrons originally was based on the
premise that there is a color spin-spin interaction between the
quarks~\cite{De Rujula:1975ge}, which is modeled after the interaction between
magnetic dipoles in electromagnetism, the so-called ``Fermi-Breit'' interaction~\cite{hyperfine}.
A few years later Isgur, Karl, and Koniuk wrote a seminal
paper~\cite{Isgur:1981yz}, 
which offered an impressive list of indirect
empirical evidence for this hypothesis.  However, in a remarkable
similarity to the flawed trinucleon deformation hypothesis of
Schwinger, due to the oversimplified description of the system, the
non-relativistic shell model description of baryons (``tri
quarks'') is now also found to be unable to quantitatively describe the deformation
when solely invoking the color magnetic tensor interaction. The inadequacy of the
non-relativistic description and of the phenomenological description
of the constituents used (lack of mesonic degrees of freedom) are
understood to be the principal deficiencies of this model.

In their paper, Isgur, Karl, and Koniuk singled out the quadrupole amplitude in
the $\Delta \rightarrow N \gamma$ transition as being a most
sensitive test of this hypothesis.  Of additional interest are the
quark model calculations, which showed that the D-state admixtures
caused by the color hyperfine interaction predict a non-zero neutron
charge distribution and  root-mean-square (RMS) charge
radius~\cite{Carlitz:1977bd, Isgur:1977mx, Isgur:1980hh}.
These theoretical speculations induced concerted experimental and
theoretical efforts to measure and calculate deviations from
spherical symmetry (non-spherical amplitudes) in hadrons.

\subsection{Size and Shape in Classical and  Non-Relativistic Quantum Mechanics}

The concepts of both size and shape, because of their familiarity in
everyday language, are often taken to be intuitively apparent, at least in
classical physics.  However, a careful examination reveals that this is not at all the case except for rigid objects with uniform
density and sharp boundaries. The size of a hurricane or the size
and the shape of nebula (e.g. the crab nebula) are not easy to
quantify.  However, the distribution in space and time of
some extensive property of an object such as its mass or charge can uniquely and
unambiguously be defined.
Its mass density $\rho(\textbf{r})$ is uniquely defined
and so is its variation in time $\rho(\textbf{r}, t)$. In
classical physics densities can be precisely defined and measured
and their knowledge allows one to define a ``size'' and a ``shape''.
Moments of the density  distribution are often quoted, which, in
simple geometrical limiting cases, have the expected correspondence to the naive concept of size or shape.
For instance the second moment of the density distribution
\begin{equation}
\langle r^2 \rangle = \int{d\textbf{r}} \cdot r^2 \cdot
\rho(\textbf{r})
\label{eq:rms}
\end{equation}
corresponds to the radius of a spherical body with
uniform distribution $\rho(\textbf{r}) = \rho_0 \theta(R-r)$.

For objects whose density distribution deviates from spherical
symmetry it is obvious that higher moments will assume non-vanishing
values. The first such moment whose non vanishing value indicates
non-sphericity is the quadrupole moment $Q_{ij}$~:
\begin{equation}
Q_{ij}  = \langle Q_{ij} \rangle = \int{d\textbf{r}} \cdot (3r_i r_j
-r^2 \delta_{ij}) \cdot \rho(\textbf{r}),
\label{eq:Qmoment}
\end{equation}
with $i, j = 1,2, 3$ denoting the spatial directions.
It is worth noting that it is possible to have non-spherical
distributions that have vanishing quadrupole moments.

 The introduction of non-relativistic quantum mechanics and the implications of the uncertainty principle have
influenced profoundly our understanding of the concept of
size. Early on, with the aid of the ``correspondence principle'' it
was realized that  "size" could be expressed in terms of the RMS radius
given by Eq.~(\ref{eq:rms}) where
$\rho(\textbf{r}) = \psi^*(\textbf{r}) \psi(\textbf{r})$ is the
probability density expressed in terms of the wave function
$\psi(\textbf{r})$ of the object. Likewise,  the quadrupole moment of
a system $\langle Q_{ij} \rangle $ manifests deviation
of its probability density from spherical symmetry.

\subsection{Size and shape in relativistic systems}

The introduction of relativity does complicate matters. It
is well understood that both the size and the shape of an object, are not
relativistically invariant quantities: observers in different frames will infer different magnitudes
for these quantities.
Furthermore when special relativity is written in a covariant formulation, the density appears
as the time (zeroth) component of a four-current  density
$J^{\mu}=(\rho,\textbf{J})$ (in units where the speed of
light $c = 1$).

\begin{figure}[h]\vspace{-1cm}
\begin{minipage}{0.49\linewidth}
\hspace*{-1.7cm}\includegraphics[width =2\linewidth]{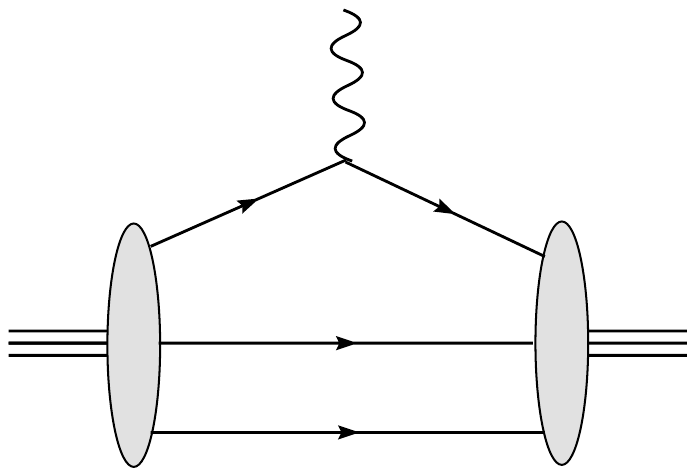}
\end{minipage}\hfill
\begin{minipage}{0.49\linewidth}
\hspace*{-1.7cm}\includegraphics[width =2\linewidth]{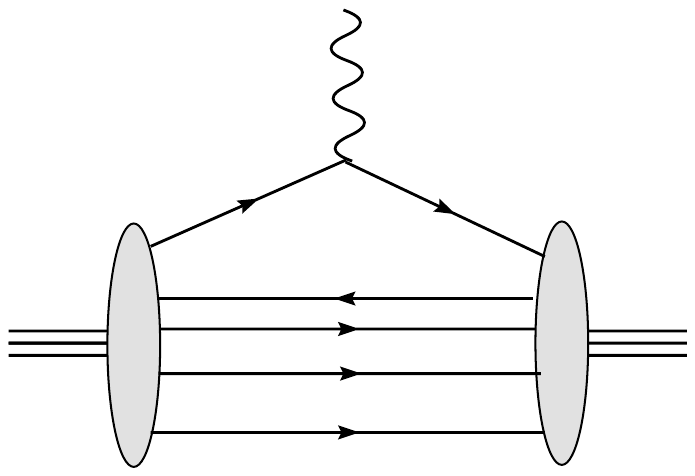}
\end{minipage}
\begin{minipage}{\linewidth}\vspace*{-9cm}
\hspace*{1cm}
\includegraphics[width=\linewidth]{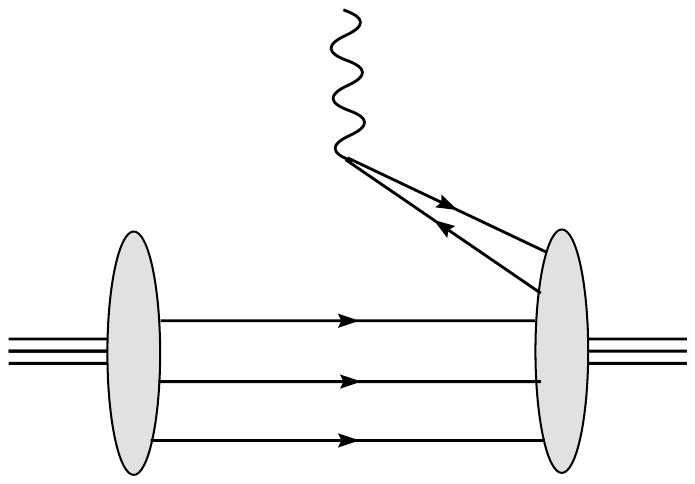}
\end{minipage}
\vspace*{-8cm}
\caption{Coupling of a space-like virtual photon to a relativistic
many-body system, as a proton. Upper panel : diagonal
transition where the photon couples to a quark, in the leading $3q$
Fock component (left), or in a higher $5q$ Fock  component
(right). Lower panel : process where the
photon creates a $q \bar q$ pair leading to a non-diagonal
transition between an initial $3q$ state and a final $5q$ state in
the proton. }
\label{fig:overlap}
\end{figure}

 Besides the relativistic kinematical effects, e.g. due to length contraction,
 the concept of size and shape in relativistic quantum systems, such as hadrons, is also profoundly
 modified as the number of constituents
is not constant as a result of virtual pair production. Consider, as an example, a hadron such as the proton, which is probed by a space-like virtual photon, as shown in
 Fig.~\ref{fig:overlap}.
 A relativistic bound state is made up of almost massless quarks. The three valence
 quarks, which make up for the proton quantum numbers,  constitute only a few percent of the total proton mass.
 In such a system, the  wave function contains,
 besides the three valence quark Fock component  $| qqq \rangle$,  also components where
 additional $q \bar q$ pairs, so-called sea-quarks, or (transverse) gluons $g$ are excited,
 leading to an infinite tower of $ |qqq q \bar q \rangle$, $ |qqq g \rangle$, ... components.
 When probing such a system using electron scattering,
 the exchanged virtual photon will couple to any quark, both valence
 and sea in the proton as shown in Fig.~\ref{fig:overlap} (upper panel).
In addition, the virtual photon,
 can also produce a $q \bar q$ pair, giving rise e.g. to a transition from a $3q$ state in the initial wave function to a $5q$ state in the final wave function, as shown in Fig.~\ref{fig:overlap} (lower panel). Such processes,
 leading to non-diagonal overlaps between initial and final wave functions, are not positive definite, and do not allow
 for a simple probability interpretation of the density $\rho$ anymore. Only the processes shown in the upper panel of
 Fig.~\ref{fig:overlap}  with the same initial and final wave function yield a positive definite particle density, allowing for a probability interpretation.

This relativistic dynamical effect of pair creation or annihilation
fundamentally hampers the interpretation of density and
any discussion of size and shape of a relativistic quantum system.
An interpretation in terms of the concept of a density requires
suppressing the contributions
shown in the lower panel of Fig.~\ref{fig:overlap}. This is possible when viewing
the hadron from a light front, which allows to describe the hadron
state by an infinite tower of light-front wave functions. Consider the electromagnetic
(e.m.) transition from an initial hadron (with four-momentum $p$) to
a final hadron (with four-momentum $p^\prime$) when viewed from a
light front moving towards the hadron. Equivalently, this
corresponds with a frame where the hadrons have a large
momentum component along the $z$-axis chosen along the direction of
the hadrons average momentum $P = (p + p^\prime)/2$.
One then defines the light-front plus  (+) component by $a^\pm \equiv
a^0 \pm a^3$, which is always a positive quantity for the quark or
anti-quark four-momenta in the hadron. When we now view the hadron in a so-called
Drell-Yan frame~\cite{Drell:1969km}, where the virtual photon
four-momentum $q$ satisfies $q^+ = 0$, energy-momentum conservation
will forbid processes where this virtual photon splits into a $q
\bar q$ pair.
Such a choice is possible for a space-like virtual photon, and its
virtuality is then given by  $q^2 = - {\vec q_\perp}^{\, 2} \equiv - Q^2 < 0$,
where $\vec q_\perp$is the transverse photon momentum
(lying in the $xy$-pane).
In such a frame, the virtual photon only couples to
forward moving partons, i.e. only processes such as those shown in the
upper panel in
Fig.~\ref{fig:overlap}  are allowed. We can then define
a proper density operator through the + component of the four-current by
$J^+ = J^0 + J^3$~\cite{Susskind:1968zz}.
For quarks it is given by
\begin{equation}
J^+ = \bar q \gamma^+ q = 2 q_+^\dagger q_+, \quad 
\mathrm{with} \quad q_+ \equiv (1/4) \gamma^- \gamma^+
q, \label{eq:reloperator}
\end{equation}
where the $q_+$ fields are related with the quark fields $q$ through a field redefinition, 
involving the $\pm$ components of the Dirac $\gamma$-matrices.
The relativistic density operator $J^+$, defined in  Eq.~(\ref{eq:reloperator}), is a positive definite quantity. For
systems consisting of $u$ and $d$ quarks, multiplying
this current with the quark charges yields a quark charge
density operator given by~$J^+(0) = +\frac{2}{3} \, \bar u(0) \gamma^+ u(0) -
\frac{1}{3} \, \bar d(0) \gamma^+(0) d(0)$. Using such quark charge density
operator, one can then define quark (transverse) charge densities in a
hadron as~\cite{Burkardt:2000za, Miller:2007uy}~:
\begin{eqnarray}
\label{eq:dens1}
\rho_\lambda(b) &\equiv& \int \frac{d^2 \vec q_\perp}{(2
\pi)^2} \, e^{- i \, \vec q_\perp \cdot \vec b} \, \frac{1}{2 P^+}
\\
&& \hspace{1.25cm}\times \langle P^+, \frac{\vec q_\perp}{2},
\lambda \,|\, J^+(0) \,|\, P^+, -\frac{\vec q_\perp}{2}, \lambda
\rangle,
\nonumber
\end{eqnarray}
with $\lambda$ the hadron (light-front) helicity.
In the two-dimensional Fourier transform of Eq.~(\ref{eq:dens1}), the two-dimensional vector
$\vec b$ denotes the quark position (in the $xy$-plane) from the
transverse center-of-momentum (c.m.) of the hadron. It is the position
variable conjugate to the hadron relative transverse momentum $\vec q_\perp$.
The quantity $\rho_\lambda(b)$ has the interpretation of the two-dimensional (transverse) 
charge density at distance $b = | \vec b|$ from the transverse c.m. of the hadron with
helicity $\lambda$. In the light-front frame, it corresponds with the projection of the
charge density along the line-of-sight.

\begin{figure}[h]
\includegraphics[width =3cm]{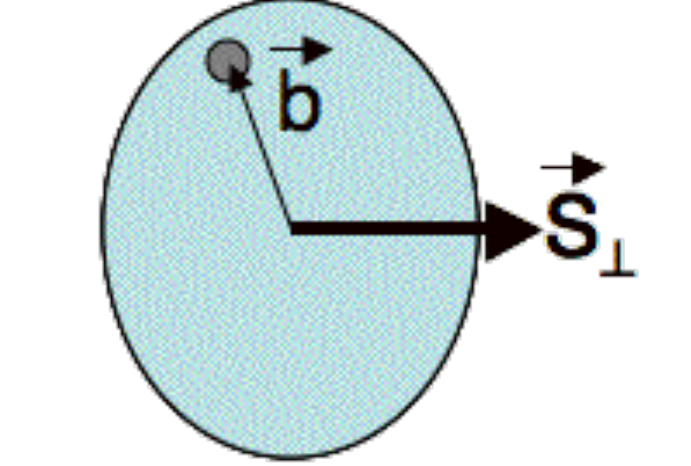}
\caption{Schematic view of the projection of the charge density along the line-of-sight (perpendicular to the figure), for a hadron polarized along the direction of $\vec S_\perp$. The position of the (quark) charge inside the hadron is denoted by $\vec  b$. }
\label{fig:relativity}
\end{figure}

The quark charge densities in Eq.~(\ref{eq:dens1})
do not fully describe the e.m. structure of the hadron, e.g. for spin 1/2
the densities with $\lambda = \pm 1/2$ yield the same
information. 
We do know however that there are two independent e.m.
FFs describing the structure of the nucleon. In general, a
particle of spin $S$ is described by $(2 S + 1)$ e.m. moments. To
fully describe the structure of a hadron one also needs to consider
the charge densities in a transversely polarized hadron state,
denoting the transverse polarization direction by
$\vec S_\perp$.
The transverse charge densities can be defined through matrix
elements of the density operator $J^+$ in eigenstates of transverse spin~\cite{Carlson:2007xd,Carlson:2008zc,Lorce:2009bs} as~:
\begin{eqnarray}
\label{eq:dens2}
\rho_{T \, s_\perp} (\vec b) &\equiv& \int \frac{d^2 \vec
q_\perp}{(2 \pi)^2} \, e^{- i \, \vec q_\perp \cdot \vec b} \,
\frac{1}{2 P^+}  \\
&& \hspace{1.25cm} \times \langle P^+, \frac{\vec q_\perp}{2},
s_\perp | J^+ | P^+, \frac{- \vec q_\perp}{2}, s_\perp  \rangle,
\nonumber
\end{eqnarray}
where $s_\perp$ is the hadron spin projection along the direction of
$\vec S_\perp$. Whereas the density $\rho_\lambda$ for a hadron
in a state of definite helicity is circularly symmetric for all spins,
the density $\rho_{T \, s_\perp}$ depends also on the
orientation of the position vector $\vec b$, relative to the
transverse spin vector $\vec S_\perp$, as illustrated in Fig.~\ref{fig:relativity}. 
Therefore, it contains 
information on the hadron shape, projected on a
plane perpendicular to the line-of-sight. 
The matrix elements of the density operator can be written in terms of
helicity amplitudes which in turn can be expressed in terms of the form factors.
From $\rho_{T \, s_\perp}$, one can then straightforwardly define e.m.
moments quantifying the shape. As an example,
for a hadron with  spin $S>1/2$,  and with transverse spin orientation
$\vec S_\perp = \hat e_x$, the electric quadrupole moment is given by~:
\begin{equation}
Q_{s_\perp} \equiv e \int d^2 \vec b \, (b_x^2 - b_y^2) \,
\rho_{T \,  s_\perp} (\vec b).
\label{eq:dens3}
\end{equation}

These light-front densities require us to develop some new intuition, as they 
are defined at equal light-front time ($x^+ = 0$) of their constituents. When constituents move non-relativistically, it does not make a difference whether they are observed
at equal time ($t = 0$) or equal light-front time ($x^+ = 0$),
since the constituents can only move a negligible small distance during the small time interval that a light-ray needs to connect them.
 This is not the case, however,  for bound systems of relativistic constituents
such as hadrons~\cite{Jarvinen:2004pi, Hoyer:2009ep}. 
For the latter, the transverse density at equal light-front time can be interpreted as a 2-dimensional 
photograph of a 3-dimensional object, reflecting the position of charged constituents at different times, which can be (causally) connected by a light ray.


\section{Measuring and Calculating the Shape of Hadrons}
\label{sec:tools}

The determination of the shape of hadrons, interesting as it may be,
presents a particularly difficult situation both theoretically and
experimentally. The challenge lies in identifying the
observables that can provide a characteristic signal, which can be
experimentally accessed with sufficient accuracy
and can be interpreted reliably to extract the information about  shape.
 This has proved to be a particularly hard
task for a number of reasons, which are discussed in this section.

It has been possible in the last decade to reach the appropriate
sensitivity and technical maturity to obtain and analyze the data
that can provide the first convincing information on the
shape of hadrons. 
To interpret the data in terms of hadronic structure quantities 
requires a reliable reaction framework. 
Such a reaction framework, as well as the interpretation and its connection to QCD,
primarily through lattice gauge calculations, have advanced to
maturity in recent years.


In this section we review and present
these advances, the experimental methods, and the theoretical
framework, which have allowed the first determination on
the shape of hadrons.

\subsection{Empirical information for spin-1 particles: $W$ boson and deuteron}
\label{sec:exp}

We start by discussing the empirical  information on the e.m.
moments of spin-1 particles, which are the  particles with the smallest spin
where a quadrupole moment can be measured. In nature, charged spin-1
particles include the $W$ gauge bosons in the Standard Model of
particle physics, the vector mesons in hadronic physics and  the deuteron 
in nuclear physics.
 For a spin-1
system, it is customary to denote the three elastic e.m. form factors (FFs)
as measured in elastic electron scattering 
 by $G_C$ (Coulomb monopole), $G_M$ (magnetic dipole), and $G_Q$ (Coulomb quadrupole), 
 where the multipole nomenclature refers to a Breit frame interpretation. 

From the empirical knowledge of the spin-1 FFs, one can map out the
charge densities in a spin-1 particle of transverse polarization by
working out the Fourier transform in Eq.~(\ref{eq:dens2}), which
yields monopole, dipole and quadrupole field patterns in the
charge density~\cite{Carlson:2008zc}. The monopole field pattern
corresponds to a circularly symmetric two-dimensional distribution for a spin-1
particle of fixed helicity. The dipole field pattern in the charge
distribution is specific for a relativistic theory.  Indeed, a
magnetic dipole moment in a rest frame manifests itself as an electric dipole
moment when seen by a moving observer, proportional to the vector
product (velocity) $\ \times$ (magnetic moment).
The induced electric dipole moment
(EDM) corresponding to the transverse charge densities $\rho_{T \,
s_\perp}$ of Eq.~(\ref{eq:dens2}) for
transverse spin projections $s_\perp = 0, \pm 1$ is given by~:
\begin{eqnarray}
\vec d_{s_\perp} &\equiv& e \int d^2 \vec b \, \vec b \, \rho_{T \,
s_\perp}(\vec b) .
\end{eqnarray}
For example, when the transverse spin projection $s_\perp =
1$, the expression for the electric dipole moment is~\cite{Carlson:2008zc}
\begin{eqnarray}
\vec d_1 = - \left( \vec S_\perp \times \hat e_z \right) \, \left[
G_M(0) - 2 \right] \, e/(2 M), \label{eq:edm}
\end{eqnarray}
where $M$ is the mass of the particle. Expressing the spin-1
magnetic moment in terms of the $g$-factor, {\it i.e.} $G_M(0) = g$,
one sees that the induced EDM $\vec d_1$ is proportional to $g - 2$.
The same result was found for the case of a spin-1/2
particle~\cite{Carlson:2007xd}. One thus observes that for a
particle without internal structure, corresponding with $g =
2$ at tree level~\cite{Ferrara:1992yc}, there is no induced EDM.

The electric quadrupole field pattern in the transverse
charge density $\rho_{T \, s_\perp}$ yields a quadrupole moment,
which is obtained,  for $s_\perp = 1$, from Eq.~(\ref{eq:dens3}) as~\cite{Carlson:2008zc}~:
\begin{eqnarray}
Q_{1} &=& (1/2)
 \left [ \left( G_M(0) - 2 \right) +
\left( G_Q(0) + 1 \right) \right] ( e / M^2 ) .
\label{eq:quadrup}
\end{eqnarray}
For a charged spin-1 particle without internal structure,
exemplified by the $W$ gauge bosons of the standard electroweak theory,
it is required that at tree level $G_M(0) = 2$ and $G_Q(0) = -1$.
For elementary particles, any deviations at tree level from these values would
indicate new, beyond standard model,  physics, and will show up in
the presence of anomalous $WW\gamma$ couplings, usually
parametrized in terms of two new couplings $\kappa_\gamma$ and
$\lambda_\gamma$, appearing in an effective Lagrangian.
In terms of those parameters, the $W$ magnetic
dipole and quadrupole moments take on the values~\cite{Hagiwara:1986vm}~:
\begin{eqnarray}
\mu_W &=& e / (2 M_W) \left [ 2 + (\kappa_\gamma - 1) + \lambda_\gamma \right ] \\
Q_W &=& - e/M_W^2 \left [ 1 + (\kappa_\gamma - 1) -\lambda_\gamma \right ] 
\end{eqnarray}
with $M_W$ the $W$-boson mass. The Standard Model values $G_M(0) = 2$
and $G_Q(0) = -1$ equivalently correspond with $\kappa_\gamma = 1$,
$\lambda_\gamma = 0$ at tree level.
The measurement of the gauge boson couplings and the search for
possible anomalous contributions due to the effects of new, beyond Standard Model,
physics have been among the principal physics aims at LEP-II.
They have been prominently studied in the $e^+ e^- \to W^+ W^-$ process through an s-channel virtual photon exchange mechanism. The most recent Particle
Data Group (PDG) fit for
the anomalous $WW \gamma$ couplings based on an analysis of all LEP
data is given by~\cite{Nakamura:2010zzi} :
\begin{eqnarray}
\kappa_\gamma = 0.973^{+ 0.044}_{ - 0.045} \quad \quad
\lambda_\gamma = -0.028^{+ 0.020}_{ - 0.021} .
\end{eqnarray}
One thus sees that present day information shows no evidence for
anomalous $WW  \gamma$ couplings, confirming the point particle
values $G_M(0) = 2$ and $G_Q(0) = -1$ for the $W$ bosons, leading to
vanishing induced electric dipole and quadrupole moments according
to Eqs.~(\ref{eq:edm}, \ref{eq:quadrup}). It is thus interesting to
observe from Eq.~(\ref{eq:quadrup}) that $Q_{s_\perp}$ is only
sensitive to the anomalous parts of the spin-1 magnetic dipole and
electric quadrupole moments, and vanishes for a particle without
internal structure.

For composite particles, it is the deviation from these benchmark
values that indicate deformations of the states. A well studied
example of a nuclear state is the deuteron. Its magnetic dipole
moment is given by $G^d_M(0) = 1.71$~\cite{Mohr:2008fa}, close to a
spin-1 particle's natural (i.e point-like) value. However, in contrast to the $W$
gauge bosons, the deuteron has a large anomalous quadrupole moment.
Its measured value is $G^d_Q(0) = 25.84 \pm
0.03$~\cite{Ericson:1982ei}. Its large value was interpreted to arise from
the prominent role of the one-pion exchange tensor interaction. One also sees from Eq.~(\ref{eq:quadrup}) that  the natural value $G_Q(0) = -1$,
arising in a relativistic quantum field theory for a spin-1 point particle,
only amounts to a few percent of the deuteron's  total quadrupole moment.
For an understanding of its static properties, the deuteron
can therefore be considered, to a good approximation, as a non-relativistic bound state system.

\begin{figure}[tc]\vspace*{-3cm}
\includegraphics[width =7.25cm]{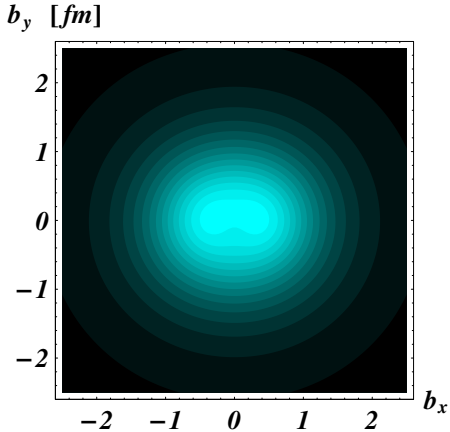}\vspace*{-0.5cm}
\caption{Two-dimensional charge density $\rho_{T \, 1}$, according to Eq.~(\ref{eq:dens2}), 
for a deuteron polarized along the positive $x$-axis. The
light (dark) regions correspond with largest (smallest) densities.
The density is calculated from empirical information for the
deuteron e.m. FFs~\cite{Abbott:2000ak}. Figure from
Ref.~\cite{Carlson:2008zc}.
\label{fig:dtrans}
}
\vspace*{-0.5cm}
\end{figure}

In the case of the deuteron its three e.m. FFs have been separated
experimentally~\cite{Abbott:2000ak}, and it has been possible to
determine the empirical charge densities. A pictorial result for
the transverse charge density with transverse deuteron polarization
$s_\perp = 1$ is shown in Fig.~\ref{fig:dtrans}. The quadrupole field pattern clearly displays a deformation along the axis of the spin ($x$-axis) together with a small overall shift of the charge distribution along the $y$-axis.

\subsection{ Measuring the shape of hadrons}

After the discussion of these two extreme cases, namely,
on the one hand of a spin-1 point particle within relativistic
 quantum field theory, and on the other hand of a non-relativistic
 two-body system, we now turn our discussion to hadrons,
such as mesons and baryons composed of light quarks.

\begin{figure}[h]\vspace*{-0.5cm}
  \hspace*{-2cm} \includegraphics[width = 0.9\linewidth]{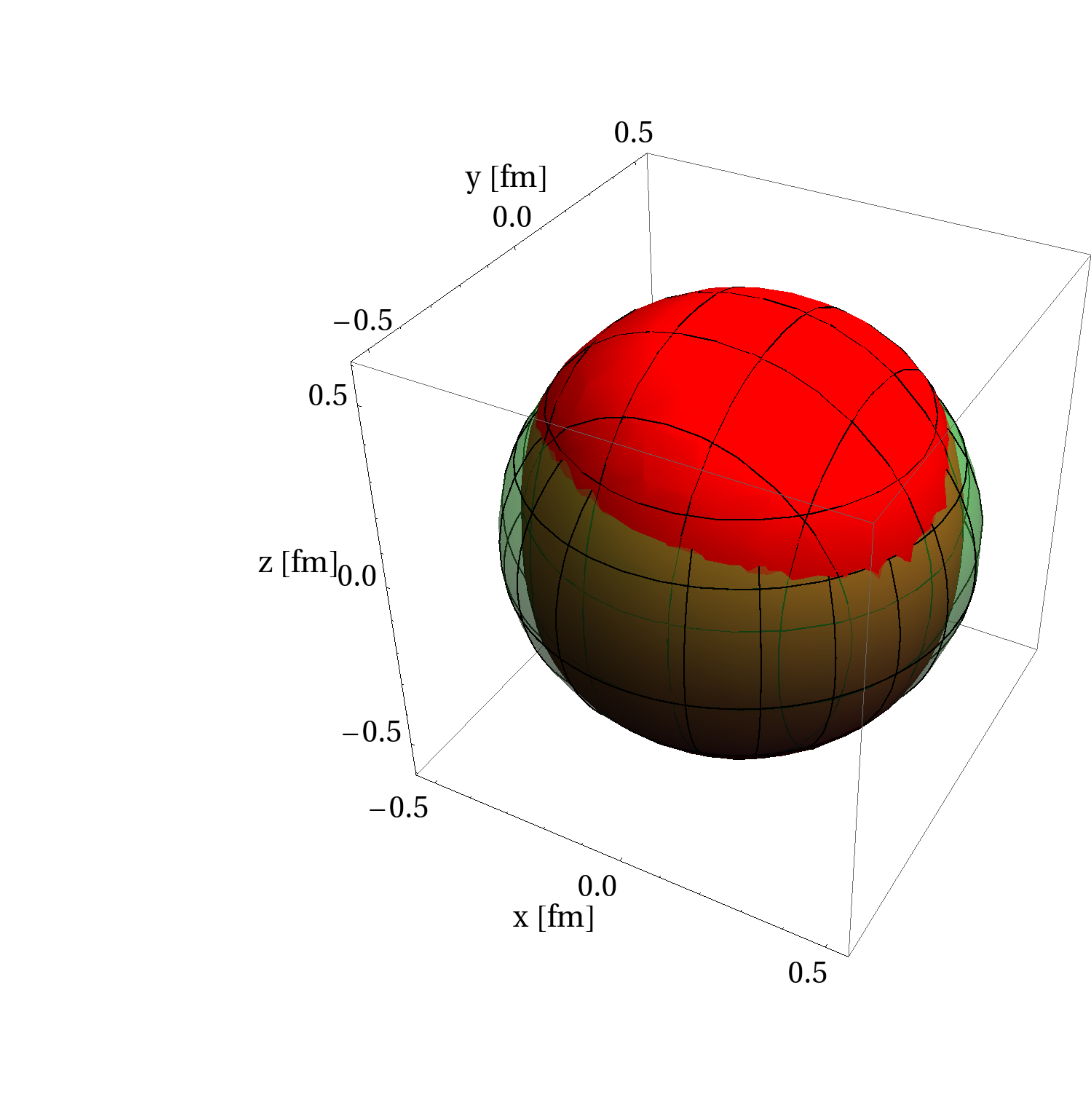}
\vspace*{-1cm}
 \caption{Three-dimensional contour plot of the $\rho$-meson, of spin projection $s_z=0$,
density-density correlator (red surface), showing
 all positions where the correlator is reduced to half its value at the origin. As can be seen, by comparing to a 
 sphere (green transparent surface) of  radius of approximately 0.5~fm, 
the $\rho-$meson surface extends outside the sphere at the poles whereas
at the equator is inside the sphere, showing the cigar-like $\rho$-meson shape. }
  \label{fig_rho_lattice}
\end{figure}

Experimentally, accessing information  that reveals
 hadron shape, even at the very rudimentary level that attempts only to check
deviations from spherical symmetry, has proved very difficult for a number of
reasons.
There is only one stable hadron, the proton, and for this reason
it is the only hadron that can provide a thick target for high
luminosity precision measurements. The relatively long lived neutron
either free or inside nuclei could provide a possible, but
technologically far more difficult alternative. Its shape
has not been explored so far.
 Both the proton and the neutron are unfortunately spin 1/2 systems and
therefore cannot provide information about their intrinsic shape
through the measurement of a static quadrupole or higher
multipole moments. From the decuplet spin 3/2 baryons  only in the case of
the $\Delta$ and the $\Omega^-$   it is
possible, in principle, to measure their quadrupole moments or the
transition quadrupole moments to some other state.
The $\Delta^+(1232)$ offers the most accessible case; however its
exceedingly short lifetime 
prevents a viable, yet,
experimental way to access its quadrupole moment. Nevertheless the magnetic
moments of the $\Delta^+$~\cite{Kotulla:2003pm} and
$\Delta^{++}$~\cite{Nakamura:2010zzi} have been
measured, albeit with very large errors. New experiments at MAMI are
expected to yield a more precise measurement for the $\Delta^+$ dipole moment.
The dipole moment of the $\Omega^-$ is more precisely measured and
provides a benchmark for lattice QCD calculations~\cite{Alexandrou:2009rk}, which in
turn   can predict its
quadrupole moment.
Vector mesons have a static quadrupole moment,
which, if different from its natural value of -1, is a clear indication of a deviation from
spherical symmetry. The $\rho$-meson is the lowest lying
spin-1 resonance to test the deviation from spherical symmetry.
However,  experimentally it is again not feasible to measure.
 A beautiful example of what information lattice QCD can
yield on hadron shapes is given in
Fig.~\ref{fig_rho_lattice}, which shows lattice calculations for the
density-density correlator of the $\rho$-meson in the lab
frame~\cite{Alexandrou:2008ru}. In the spin projection
zero case, the $\rho$-meson displays a prolate (cigar-like) deformation in its
rest frame. This conclusion is corroborated by a  calculation of
the $\rho$-meson quadrupole moment in quenched lattice QCD~\cite{Hedditch:2007ex}.

Thus to measure the shape of hadrons, 
none of the ``standard" and tested methods used in atomic and nuclear
physics can be employed.
The only viable path to study the nucleon shape
 remains the one originally proposed by Isgur and Karl,
i.e. to measure the presence of resonant
quadrupole admixtures in the $\gamma^* N\rightarrow \Delta$
transition, which amounts to determining the
off-diagonal (transition) quadrupole moment. The theoretical
framework of interpreting these measurements has matured in recent
years, as will be reviewed below. The precision
measurements of this transition provide the most reliable
information we have today for deviation from spherical shape for the
proton and/or the $\Delta^+(1232)$~\cite{Papanicolas:2007zz}.

\begin{figure}[btc]\vspace*{-3.5cm}
\includegraphics[width=0.8\linewidth]{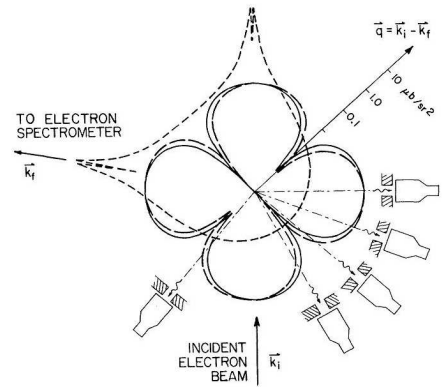}
\caption {The detection of the de-excitation radiation pattern from
a system allows to isolate the contributing multipoles. The
isolation of the multipole FFs was achieved for the first
time in the $^{12}C (e,e'\gamma)$  and $^{15}N (e,e'\gamma)$
reactions where the E2 and M1 FFs were isolated through the
tagging of the decay radiation~\cite{Papanicolas:1985zz}. The
figure shows the radiation pattern and the isolated FFs
for this transition in $^{15}N$.}
\label{fig:eegamma}
\end{figure}

The experimental technique employed in the determination of the deviation from sphericity in
the study of the de-excitation of the $\Delta^+(1232)$ resonance is
different than those discussed earlier. It involves the detection of the radiation pattern of the emitted radiation in the de- excitation of the excited state.
The concept behind the technique derives from classical electromagnetism. The
observed radiation pattern, its multipole content to be precise, reveals information about the
shape of the radiating antenna. The radiation  emitted in the de-excitation of the the $\Delta^+(1232)$
is primarily in the form of pions but a small (0.7\%) branch of $\gamma$ rays is also present.
This technique of measuring shape rarely has been used in nuclear physics, principally due to the experimental
complexity it presents.  An important exemption is the study of the $^{15}N$ excited states
using this technique in an $(e,e' \gamma)$ experiment, which demonstrated both the feasibility and accuracy of
this method. The 6.33 MeV $J^{P}=3/2^{-}$ excited state of $^{15}N$, a $J^{P}=1/2^{-}$ nucleus,
presents a case where a transition to it from the ground state with the same quantum numbers as the $\gamma^*N\rightarrow \Delta$
can be studied.  The experiment,  where the C2 (Coulomb quadrupole), E2 (electric quadrupole)
 and M1 (magnetic dipole) FFs were isolated through the
tagging of the decay radiation~\cite{Papanicolas:1985zz}, offers a clear demonstration of the power of the technique. The experimental arrangement used is shown in
Fig.~\ref{fig:eegamma}, which clearly
portrays the concept of this experimental technique.  The virtual photon causes the excitation of the target nucleus
and due to angular momentum  and parity selection rules only magnetic dipole and electric quadrupole transitions are allowed; the decay radiation pattern allows to identify each admixture.

In the excitation spectrum of the nucleon the only isolated state is the
$\Delta^{+}(1232)$, which thus allows us to employ the same type of measurement as in the 6.33~MeV isolated excited state of $^{15}N$.  The $\gamma^{*}N \rightarrow \Delta$ transition from
$J = 1/2$ to $J = 3/2$ with no change in parity allows us to observe quadrupole E2 and C2
transition moments. It is however a mixed transition, which, in
addition to the quadrupole amplitudes, involves the M1 (spin flip)
amplitude that is the dominant one.  The presence of resonant quadrupole strength signifies deviation from sphericity of the proton and/or the $\Delta^+(1232)$.  Using the same experimental technique as
in the case of $^{15}N$ it is possible to isolate and measure the weak but
important quadrupole amplitudes in the presence of the dominant M1 transition.
Through the extensive study of the $N\rightarrow \Delta$ transition, 
 pursued  during the last thirty
years using real or virtual photon probes, an extensive body of data
 has emerged that convincingly
demonstrates that the quadrupole amplitudes are substantial and far
bigger than can be accommodated by a 'spherical' proton.

\begin{figure}[h]\vspace*{-5.5cm}
\includegraphics[width=1.7\linewidth,angle=-00]{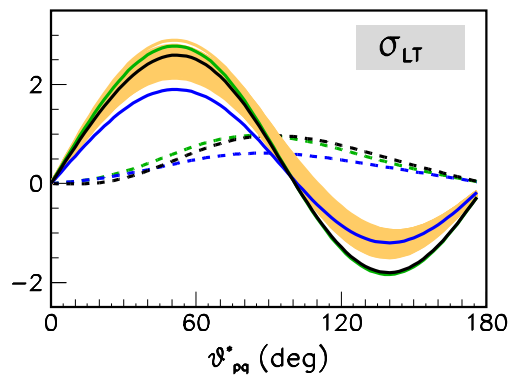}
\vspace*{-8.5cm}
\caption {The precise range of uncertainty that is allowed by the Bates and MAMI data for the
$\sigma_{LT}$ response of the $p ( e, e^\prime p) \pi^0$ reaction,
at $Q^{2}=0.127$~GeV$^2$ and at the position of the $\Delta^+(1232)$ resonance,
is shown as a function of the c.m. angle between the proton and virtual photon $\theta^*_{pq}$.
Current phenomenological models (MAID~\cite{Drechsel:2007if} in black, Sato-Lee (SL)~\cite{SL} in blue and DMT~\cite{DMT} in green)
predict this satisfactorily within a $2\sigma$ confidence level. 
The corresponding calculations for ``spherical" nucleon and $\Delta$ (dashed curves using same color coding for the various models) cannot describe the data; they
are excluded to $2~\sigma$ confidence level.}
\label{fig:errorbands}
\end{figure}

Experimentally, the measurement of the small value of the electric or Coulomb quadrupole multipole
becomes possible through its interference with the dominant magnetic dipole transition.
This is shown in Fig.~\ref{fig:errorbands} which depicts the angular dependence of the
longitudinal-transverse interference  cross section ($\sigma_{LT}$)
for the $p ( e, e^\prime p) \pi^0$ reaction on top of the $\Delta^+(1232)$ resonance.
The cross section $\sigma_{LT}$ is overwhelmingly driven by the interference of the dominant transverse M1 amplitude with the longitudinal C2 amplitude.
The experimentally constrained region, by the Bates and MAMI data, is compared
with phenomenological model predictions that attempt to
describe the experimental data. It is evident that the model
predictions (dotted curves) with resonant quadrupole amplitudes set
to zero, which amounts to spherical solutions, are excluded with
high confidence. The ``deformed" model predictions, assuming
 negligible model error, are in agreement
at the $2\sigma$ level, with the empirical results.
This comparison demonstrates that compelling experimental evidence nowadays exists
supporting the conjecture of deformed  hadrons. In particular,
 the above data 
demonstrate  with very high confidence
that spherical symmetry for both the nucleon and the
$\Delta^{+}(1232)$ is experimentally excluded.
The experimental results for hadron deformation in the $\gamma^* N \to \Delta$
transition will be discussed in more detail in Section~\ref{sec:experimental}.

\subsection{Calculating the shape of hadrons : lattice QCD\label{sec:lattice}}

Having seen clear experimental evidence for a non-spherical charge distribution in the $N \to \Delta$
transition, we next examine whether this can be calculated and understood from QCD, the underlying
theory of strong interactions.

QCD
requires a new methodology in order to evaluate quantities
 related to hadron structure, the reason being that hadrons are
 bound state systems having
a mass that is mostly generated by the interaction rather than by the
sum of the mass of their constituents.
 Perturbative QCD has been very successful in describing high
energy processes. On the other end of very low energy, chiral
perturbation theory has provided the appropriate effective field
theory
 framework for precise calculations of observables in terms of a small
expansion parameter, such as an external momentum or pion mass. This framework
provides  a
systematic expansion involving an increasingly large number of
low-energy constants (LECs). The latter are free parameters, which
are beyond the predictive power of the effective field theory. Some
of these have been determined from phenomenological information,
however, the vast majority remains unknown limiting the predictive
power of chiral effective field theory.
 To calculate LECs from the underlying theory of QCD as well as to make
predictions beyond a regime where a perturbative or small scale
expansion is applicable  requires an inherently non-perturbative
technique. Such an approach that enables us to solve the theory
in the non-perturbative domain starting from the underlying QCD Lagrangian
 is lattice QCD, a discretized
version of QCD
 formulated in terms of Feynman's path integrals
on a space-time lattice preserving
 gauge symmetry ~\cite{Wilson}.
Like the continuum theory,  the 
only  parameters are the bare quark masses and the coupling constant. One
recovers continuum physics by extrapolating results obtained at
finite lattice spacing $a$ to $a=0$.

A crucial
 step, that enables  one to numerically evaluate the path integrals
needed, is rotation to imaginary time, $t\rightarrow -it$, resulting
in the replacement of the time evolution operator $\exp(-i{\cal H}t/\hbar)$ by
$\exp(-{\cal H}t/\hbar)$. Within the Feynman path integral formulation,
 observables are  calculated by  a weighted sum over all possible trajectories.
 In imaginary time it becomes possible 
 to generate a representative ensemble of trajectories
by using stochastic methods 
analogous  to those applied  in the evaluation of observables  in  statistical mechanics.

Calculations in lattice QCD started in the early 80's,
and during  the first two decades were performed mostly in the
quenched approximation, which neglects pair creation.  This enormously
simplifies the generation of the  gauge fields via Monte Carlo methods since one is left with a
local gauge action. 
 During the past ten years, theoretical
progress in combination with terascale computers have made
simulation of the full theory  with light pions and large enough volumes
feasible using several different
discretization schemes. The simplest lattice QCD action is due to
Wilson~\cite{Wilson}.  
Nowadays,  one uses improved discretized versions of the
Dirac operator with reduced finite lattice spacing artifacts and better
chiral properties, 
 all of which are expected to
yield the same results in the continuum limit~\cite{Jansen:2008vs}. Using these
improved fermion discretization schemes,
simulations with pion masses
 within 100 MeV of  the physical pion mass are currently available
with  
simulations using improved Wilson fermions  even reaching  the 
physical pion mass~\cite{BMW}.
 A benchmark calculation for lattice QCD is the evaluation of the low lying hadron spectrum, where a systematic study of the hadron masses
using different discretization schemes has been performed and the continuum and infinite volume limits have been examined.
The agreement with experiment observed from such systematic 
lattice studies~\cite{BMW, Alexandrou:2009qu}, provides a validation of the lattice QCD 
approach,
paving the way to use lattice QCD  to provide predictions for quantities, which
are very difficult to access experimentally, as for example the
e.m. FFs of an excited hadronic state, such as
the $\Delta(1232)$ resonance. It furthermore
 allows to study how the
physics is affected when varying fundamental parameters such as
quark masses outside their values realized in nature.

\begin{figure}[h]
\scalebox{0.45}{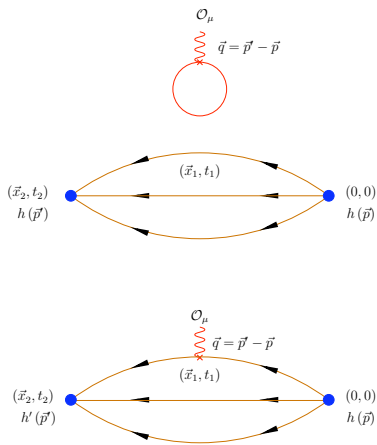}
 \caption{Upper panel: Disconnected; Lower panel: connected diagrams. With $h$ and $h^\prime$ we denote hadronic states and with ${\cal O}$  the operator of interest.}
\label{fig:diagrams}
\end{figure}

 Information on hadron shapes can be extracted from FFs
and generalized parton distributions.
The evaluation of these quantities is more involved than the computation of hadron masses.
FFs are connected to  hadron matrix elements of the type
 $\langle h^\prime(p^\prime)|{\cal O}| h(p)\rangle$ and one needs, in general,
to compute the diagrams shown in Fig.~\ref{fig:diagrams}, where the
solid lines denote fully dressed quark propagators.
The diagram where the operator couples to a sea quark, shown
in upper panel of Fig.~\ref{fig:diagrams},  is
particularly difficult to calculate since it involves a disconnected
quark loop.
For the evaluation of transition FFs where
the final hadron state $h^\prime$ has different quantum numbers
 from the initial $h$,
the disconnected diagram vanishes. For diagonal matrix
elements, assuming isospin symmetry, the disconnected contribution vanishes  for isovector
operators and therefore isovector FFs can be calculated from the connected diagram alone.
 Although recently efforts to
calculate such disconnected contributions have intensified,
 up to now lattice computations of FFs
generally neglect disconnected contributions.
 The standard procedure to evaluate the connected three-point
function shown in the lower panel of Fig.~\ref{fig:diagrams}  is  to compute the so called
sequential propagator, a convolution of fixed source quark propagators, which
are technically straight forward to calculate.
In most  recent studies 
 of the e.m. FFs the  fixed sink method is used. Its name comes from the fact
 that we consider a given final state $h^\prime$
is at a {\it fixed} time $t_2$ from the initial state $h$ created at time zero.
Within this approach, any operator
can  be inserted at  any intermediate time slice $t_1$, as seen in
Fig.~\ref{fig:diagrams}, carrying any possible value of  the lattice
momentum.  For a recent review, which includes
comparison of  nucleon electromagnetic FFs within a number of different discretization schemes   see Ref.~\cite{Hagler:2009ni}.

To probe hadron deformation, the e.m. current is used for the operator ${\cal O}$ in Fig.~\ref{fig:diagrams}.
 Lower moments of transverse spin
densities of quarks in the nucleon~\cite{Gockeler:2006zu} or
pion~\cite{Brommel:2007xd} as well as the transverse momentum dependent
parton distribution functions~\cite{Hagler:2009ni}  can also be evaluated using these
techniques. 

\begin{figure}[h]\vspace*{-3.5cm}
\begin{center}
\includegraphics[width = 7.25cm]{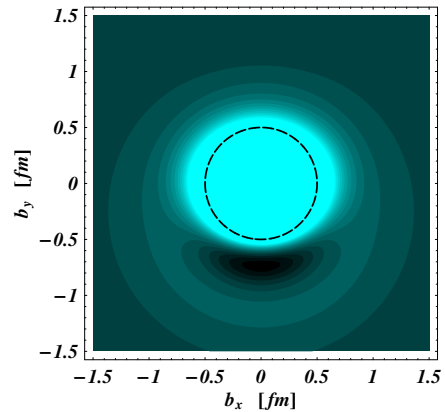}
\end{center}
\vspace{-0.5cm}
\caption{Lattice QCD results for the quark transverse charge
density $\rho^\Delta_{T \, \frac{3}{2}}$ in a
{\it $\Delta^+(1232)$} which is polarized along the positive $x$-axis.
The light (dark) regions correspond to the
largest (smallest) values of the density.
In order to see the deformation more clearly, a circle of radius 0.5~fm is drawn for
comparison.
The density is obtained from quenched lattice QCD results at $m_\pi = 410$~MeV
for the $\Delta$ e.m. FFs~\cite{Alexandrou:2009hs}.}
\label{fig:deltatrans}
\end{figure}

As an example of the predictive power of lattice QCD, we show in
Fig.~\ref{fig:deltatrans} the transverse charge density of Eq.~(\ref{eq:dens2})
for a $\Delta^+(1232)$ that has a transverse spin projection $s_\perp = +3/2$.
This charge density is obtained from the $\Delta$ e.m. FFs, calculated within lattice QCD in
Ref.~\cite{Alexandrou:2009hs}, as  described in more
detail in Section~\ref{sec:theory}.
As can be seen from Fig.~\ref{fig:deltatrans}, the  quark charge density in a
$\Delta^+$ in a state of transverse spin projection
$s_\perp = +3/2$ is elongated
along the axis of the spin (prolate deformation) when observed from a light-front.

Although lattice QCD provides an {\it ab initial} calculation of fundamental quantities
such as FFs or moments of generalized parton distributions,
a careful analysis of statistical and systematic errors  must be performed before one can reliably
compare to experiment. The systematic errors arise 
 because lattice calculations
necessarily are performed for a finite lattice size and  spacing $a$ 
due to the discretization of space-time, which 
breaks continuous rotational invariance to a discrete one.  
These systematic errors need to be investigated by repeating the calculation
for various volumes and lattice spacings.
Except for hadron masses, where both the infinite volume and zero lattice spacing limits are taken, for other quantities like FFs 
such an analysis has just began.
Another source of systematic error is the fact that FFs calculations still utilize
dynamical quarks of larger mass than the physical one.
Whereas finite $a$, lattice size $L$ and magnitude of the quark masses
 are amenable to
systematic improvements, rotation to Euclidean space selects a set
of observables,  determined from the properties of the
discrete low-lying states,
which can be  studied within this framework.
Nucleon FFs and moments of parton
distributions  are examples of such observables.
Excited states are more difficult to compute since they are  exponentially suppressed as compared to the ground state due to the Euclidean time evolution.
Techniques have been developed  to extract the low lying excited states,
 however most
calculations are still done in quenched QCD and without an analysis of systematic errors,
although some recent results on the excited states of the nucleon using two dynamical quarks have been
presented. 
The study of resonances in lattice QCD is a recent activity.  One of the reasons is that up to very recently
the quark masses that could be simulated were too large to allow decays.  Although  extraction of the spectral function from
lattice correlators
is not feasible since the low energy
continuum scattering states dominate, there are theoretical techniques to study the width of resonances~\cite{Luscher:1991cf} that make use of the dependence
of the energy on the finite lattice length.  
These techniques, combined with the background field method,
can yield the magnetic and electric  quadrupole moments of resonant states~\cite{Aubin:2008qp}. 
 However, to go beyond the calculation of the decay width and the lower moments to
the calculation of FFs for resonances such as the $\Delta$ is still an open theoretical problem. 

In Section~\ref{sec:theory}, we will present results showing the state-of-the-art of the
lattice calculations for the e.m. FFs of the $\Delta(1232)$ resonance, as well as for the e.m.
FFs describing the $\gamma^* N \to \Delta$ transition, and discuss the resulting theoretical predictions
for hadron deformation.


\section{Experimental Evidence}
\label{sec:experimental}

The experimental landscape concerning the investigation of the shape of
hadrons has been  dominated by the quest for resonant quadrupole amplitudes in the \GNdelta ~transition in the proton. Recently other reactions have been suggested, e.g. the study of the \GNdelta ~transition in  neutrons or in nuclei,
and they may become technically feasible in the near future. In addition, it is understood that the detailed and precise understanding of form factors 
can bring new complementary information on the issue of the shape of hadrons. In addition to the formidable technical difficulties of accessing new  reaction channels to the required precision, the theoretical framework to extract the important physical conclusions
needs to be further developed.

The experimental investigation of the \GNdelta~ transition
can be classified according to the reaction channel probed.
The $\Delta^{+}(1232)$ can be excited by real
or virtual photons, $\gamma^{*}$, and decays through pion or photon emission:
\begin{eqnarray}
\gamma^{*} p &\rightarrow& \Delta^{+}(1232)\rightarrow p \pi^0~~
(66\%), \nonumber \\
\gamma^{*} p &\rightarrow& \Delta^{+}(1232)\rightarrow n \pi^{+}~~
(33\%), \nonumber \\
\gamma^{*} p &\rightarrow& \Delta^{+}(1232)\rightarrow p \gamma~~~~
(0.56\%). \nonumber
\end{eqnarray}
\noindent The pion decay
channels have been extensively explored while the third, involving
the gamma decay branch, has been studied with real compton
scattering (RCS). Virtual compton scattering
measurements (VCS) are  beginning to
emerge with the aim of mapping the polarizabilities at
high missing mass~\cite{Bensafa:2006wr} and/or investigating the issue of
deformation~\cite{Sparveris:2008jx}.

The first generation  \GNdelta~and in general  nucleon resonance
 experiments were conducted in the late sixties and early seventies,
before the issue of deformation was even raised, 
 at DESY, NINA and CEA with low quality beams and
experimental equipment not designed to address such refined
questions. The data that emerged were characterized by limited accuracy,
but they did provide valuable
guidance on the design of future
experiments~\cite{CNP1989}. The second generation experiments were
obtained by a newer generation of  accelerators at Brookhaven, Bates, MAMI,
and CEBAF with optimized  equipment and in general with polarized beams.
Third generation experiments are now beginning to emerge; they have
been conducted primarily with polarized and tagged real photons,
impinging on polarized targets. Electroproduction experiments with
polarized targets are particularly difficult with only one
measurement reported in the literature using the internal target
facility at NIKHEF ~\cite{vanBuuren:2002im} and having
low statistical accuracy.
The JLab Hall A
experiment~\cite{Kelly:2005jj}, which presented high quality extensive
recoil polarization measurements using polarized beams, is a truly
third generation experiment, which both demonstrated the feasibility
of the technique, the precision that can be achieved and the rich
physics output that can emerge.

In general, in the real photon sector, the ``second generation"
experiments are completed and analyzed and the era of ``third
generation" experiments is about to begin in earnest, in view of the
important instrumentation initiatives~\cite{Kotulla:2005eg}
 at Mainz and at
Bonn. JLab and MAMI C have optimal beams and
detection systems for the pursuit of this program, which is far from
being exhausted.

\begin{figure}[ht]
\begin{center}
\includegraphics[width =0.65\linewidth,angle=-90]{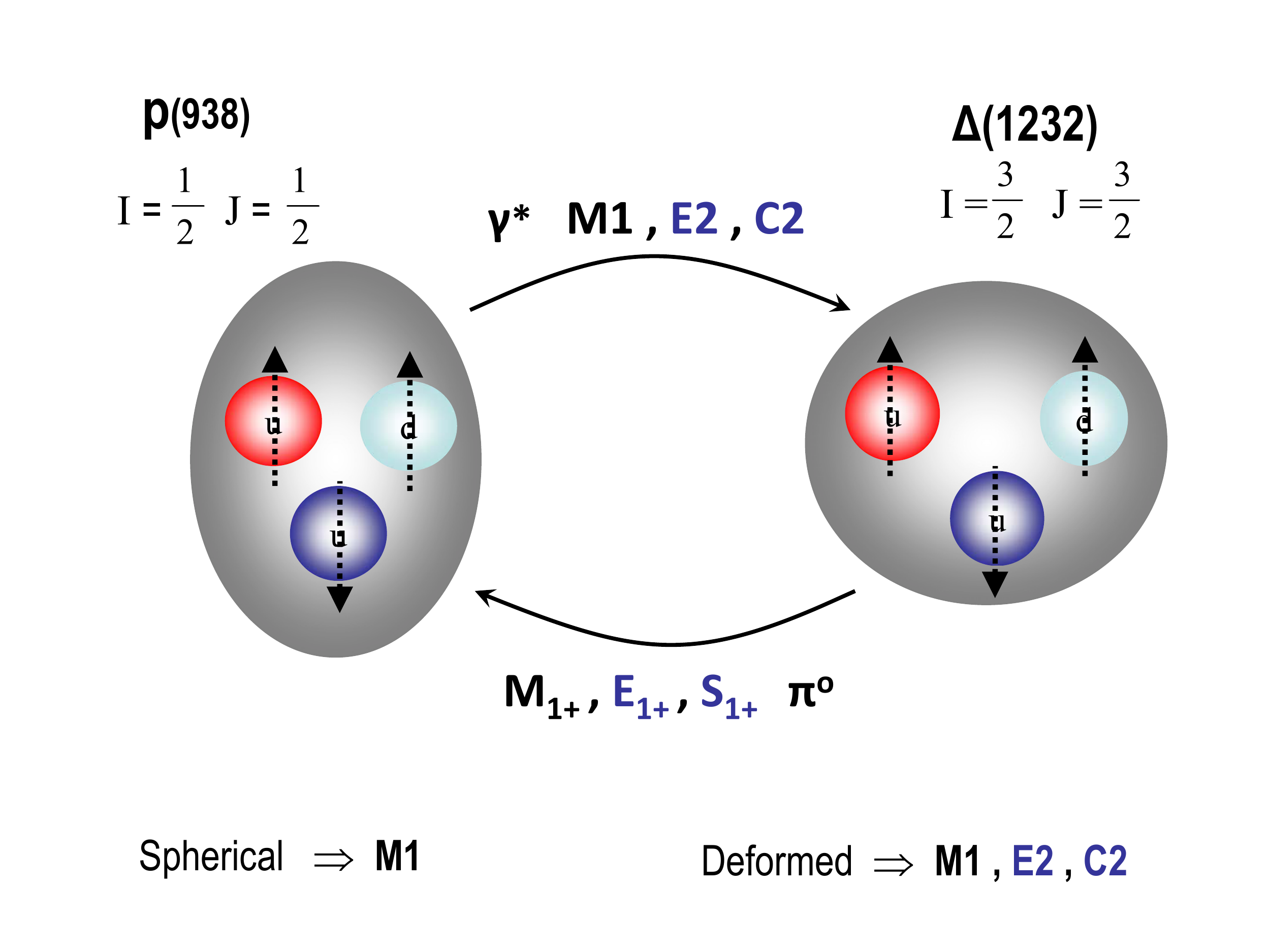}
\end{center}
\vspace*{-0.5cm}
\caption{Quark model picture of  $M1$, $E2$ and $C2$ amplitudes in the $N \to \Delta$ transition
induced by the interaction of a photon (real or virtual) with a single quark in the nucleon.
Presence of quadrupole amplitudes in the transition requires $N$ and/or $\Delta$
wave functions to have a $D$-wave component (indicated by a non-spherical shape).
}
\label{fig:m1e2ndel}
\end{figure}

The ``deformation" signal in the real photon sector comes from the
study of the  transverse electric quadrupole ($E_{1+}^{3/2}$ also denoted by E2) multipole. If virtual photons are used, the longitudinal quadrupole ($L^{3/2}_{1+}$ or C2) is also accessed. The
superscript indicates the total isospin 3/2, whereas the subscript denotes the $l = 1$ angular momentum in the $\pi N$ system, and the ``+" refers to the total angular momentum $J = l + 1/2 = 3/2$.
In a quark-model picture,  the $\gamma N \to \Delta$ transition is described by a spin flip of a quark in an $S$-wave state in the nucleon, resulting in a magnetic dipole ($M^{3/2}_{+1}$ or M1) transition. Any $D$-wave admixture in the nucleon {\it or} the $\Delta$ wave functions, also allows non-zero values for the electric quadrupole (C2 and E2) transition. This is depicted graphically in  Fig.~\ref{fig:m1e2ndel}.

\begin{figure*}[tbh]
\begin{minipage}{0.49\linewidth}
\includegraphics[width=\linewidth,angle=-00]{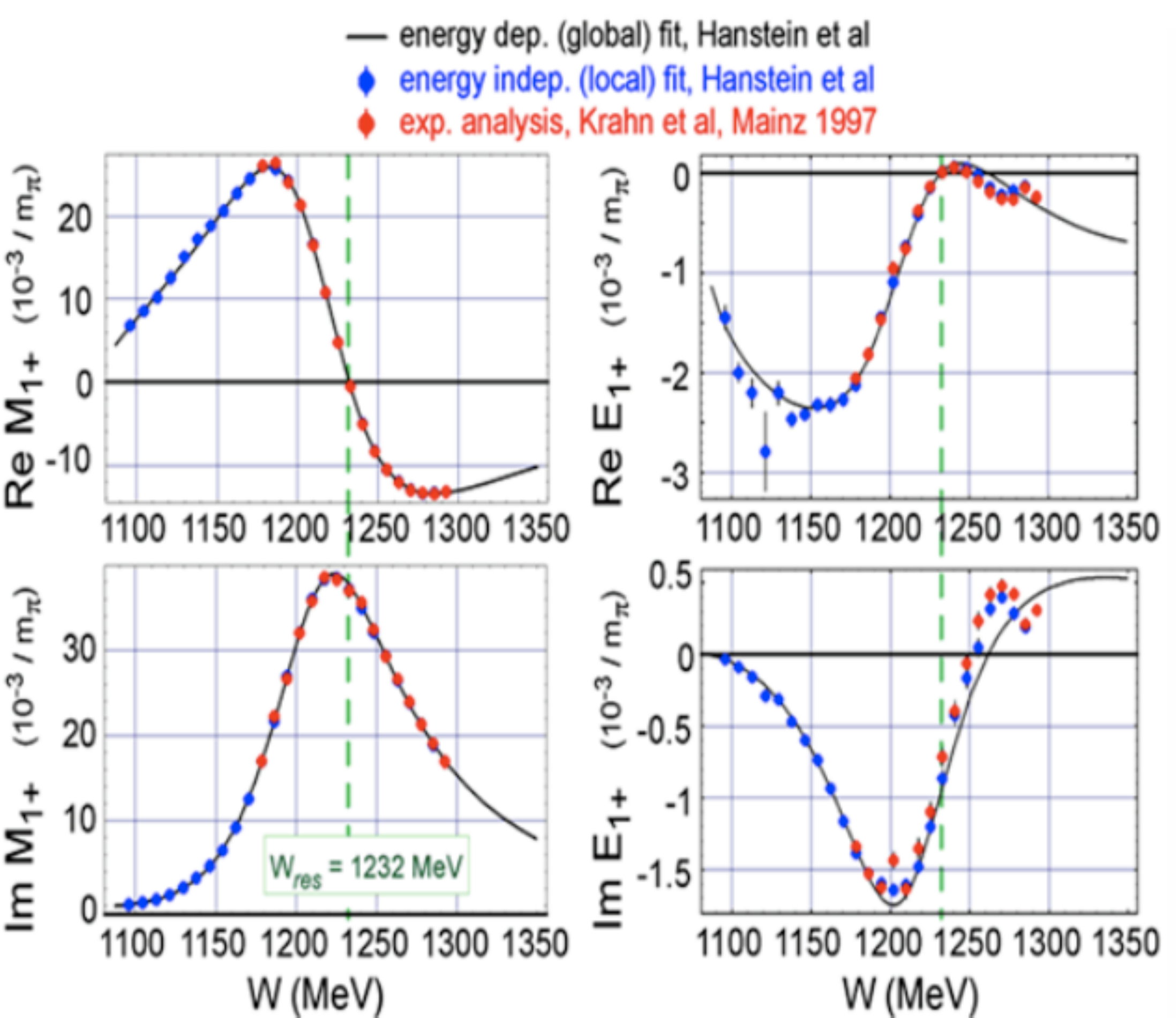}
\end{minipage}\hfill
\begin{minipage}{0.49\linewidth}
\includegraphics[width=0.7\linewidth,height=0.8\linewidth,angle=-00]{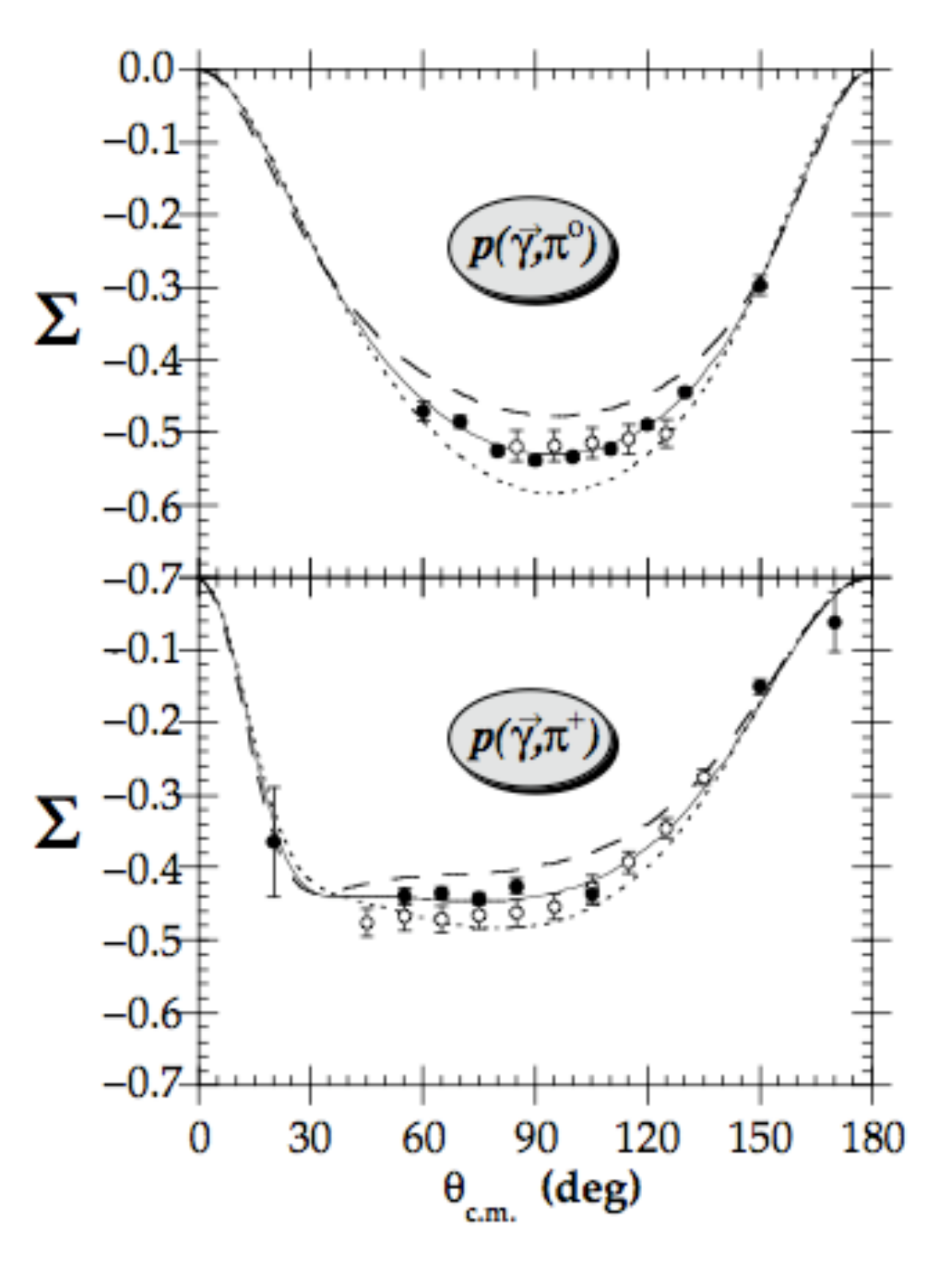}
\end{minipage}\hfill
\caption{
Results from MAMI (left) and LEGS (right) have yielded precise
measurements of the resonant quadrupole amplitude at the photon
point. A most sensitive probe is the polarization asymmetry
$\Sigma$
which has been measured precisely at both MAMI and LEGS. The derived
multipoles from the MAMI cross sections, yield an accurate
measurement of EMR.}
\label{fig_realphtons}
\end{figure*}
It has become standard practice in the field to measure the resonant
quadrupole strengths relatively to the resonant dipole by introducing the ratios
EMR$={\rm Im}E^{3/2}_{1+}/{\rm Im} M^{3/2}_{1+}$ and CMR$= {\rm Im}L^{3/2}_{1+}/{\rm Im} M^{3/2}_{1+}$. EMR and CMR have thus become the signal of deformation.

\subsection{Real Photon Measurements}
In photoproduction, the presence of a resonant quadrupole amplitude, is particularly hard to isolate because the
transverse channel is overwhelmed by the magnetic dipole (M1) amplitudes and contaminated with other "background" (non resonant) processes of similar magnitude.
In this sense, the $E_{1+}^{3/2}$ appears in next to leading order
(NLO) in photoproduction.
Precision measurements with polarized tagged photons performed at
Mainz (MAMI) and Brookhaven (LEGS) in the late nineties
represent {\it a tour de force} of experimental finesse. 
The small quadrupole amplitude has been detected in the measurement of the polarization asymmetry $\Sigma=(\sigma_{\|}-\sigma_{\bot} )/(\sigma_{\|}+\sigma_{\bot} )$  shown in the right panel of Fig.~\ref{fig_realphtons}. The asymmetry $\Sigma$  is measured with reduced systematic error by flipping the polarization of the tagged photon beam parallel $(\|)$ and perpendicular $(\bot)$ to the scattering plane.
Analysis of the MAMI $(\gamma, \pi^{+})$ and $(\gamma, \pi^0)$
data, yields the impressive results shown in the left panel of
Fig.~\ref{fig_realphtons}. It is obvious from this figure that the
derived results heavily depend on the W dependence of the cross
section. The $E_{1+}^{3/2}$
multipoles have a striking non-resonant shape, a manifestation of
the complicated processes that contribute to this channel. 
The measurements from MAMI~\cite{Beck:1997ew} and LEGS~\cite{Blanpied:1997zz}
 have converged as far as
the determination of the   asymmetries are concerned. The resulting EMR values are~:
\begin{eqnarray}
\mathrm{LEGS} &:& \mathrm{EMR} =
-( 3.07 \pm 0.26_{stat. + syst.} \pm 0.24_{mod.} ) \%,
\nonumber \\
\mathrm{MAMI} &:& \mathrm{EMR} =
-( 2.5 \pm 0.1_{stat.} \pm 0.2_{syst.} ) \%.
\nonumber
\end{eqnarray}
A number of
theoretical calculations are in good agreement with the
experimentally derived EMR value.  Both the $(\gamma, \pi^0)$ and the
$(\gamma, \pi^{+})$ channels have been studied extensively. The
$(\gamma, \gamma)$ channel (RCS) has also been studied~\cite{Galler:2001ht}, where the
resonance pion-photoproduction amplitudes were evaluated leading to
the multipole EMR (340 MeV) =$(-1.6 \pm 0.4_{(stat+syst)}
\pm 0.2_{(model)}$)\%, in reasonable agreement with the photopion
measurements.

The situation concerning the $\gamma N\rightarrow \Delta$~transition in the real photon
sector has remained stable, without experimental results reported to
change this picture in the last five years. A subsequent
analysis~\cite{Arndt:2001by} and new data~\cite{Kotulla:2007zz} give EMR = $(-2.74 \pm
0.03_{(stat)} \pm 0.3_{(syst)} )\%$, confirming the EMR values
of Ref.~\cite{Blanpied:1997zz,Beck:1997ew}.
In the closely related areas of
threshold pion production~\cite{Merkel:2006ya} and in the measurement of
the magnetic dipole~\cite{Kotulla:2003pm, Kotulla:2002cg} of the
$\Delta^{+}(1232)$, the precise results that emerged  provide
both a test as well as  valuable guidance to theory and phenomenology that is
common to both. The installation of the Crystal Ball at MAMI
and of the Crystal Barrel at ELSA, have brought  new
very powerful tools, which are expected to yield even more precise data and
results.

\subsection{Electroproduction measurements}

In electron scattering experiments, in addition to the transverse
responses, the longitudinal responses are also accessible, which are sensitive to leading order to the longitudinal quadrupole multipole, $L_{1+}^{3/2}$ or C2. Furthermore, the
$Q^{2}$ evolution of the various responses offers the ability to distinguish between different
processes. This is of particular value for the understanding of the distinctive roles played by
the mesonic cloud as compared with the quark core. These
advantages are  technically challenging and time consuming to realize due,
primarily, to the numerous measurements needed to cover the widest
possible range of momentum transfers.

Consistent results have been reported  from several groups~\cite{Kalleicher:1997qf, Warren:1999pq, Frolov:1998pw, Mertz:1999hp, Pospischil:2000ad, Kunz:2003we, Joo:2001tw, Sparveris:2004jn, Kelly:2005jj, Elsner:2005cz, Stave:2006ea, Ungaro:2006df, Sparveris:2006uk, Stave:2008tv, Kirkpatrick:2008pk} at Bates, ELSA, MAMI and JLab mapping the momentum
transfer range from $Q^{2}$= $0.06$ to $6.0$ \gevc ~with high
precision 
in a limited number of  observables sensitive to the issue of deformation. 
However, there are still  discrepancies on the extracted 
EMR and CMR  values, which are not directly measurable  due to the
methodology  used in extracting multipoles, an issue
discussed in the next section.

Starting from the experimental observables, two methods have been
used for  extracting multipole
amplitudes: a) The Truncated Multipole Expansion (TME)
approximation in which most or all of the non-resonant multipoles are
neglected (e.g. see~\cite{Kalleicher:1997qf, Frolov:1998pw}) assuming that, at
resonance, only the resonant terms contribute significantly and are fitted to the data, and b)
the Model Dependent Extraction (MDE) method where a phenomenological
reaction framework with adjustable quadrupole amplitudes is used,
e.g. see~\cite{Frolov:1998pw, Mertz:1999hp, Stave:2006ea}.
It is assumed that the reaction is controlled at the level of
precision required for the disentanglement of the background from
the resonance amplitudes. Clearly the MDE method is superior, given the
sophistication that phenomenological models have achieved in
describing the  data.

\begin{figure}[h]\vspace*{-12.5cm}
\begin{center}
\includegraphics[width =15cm]{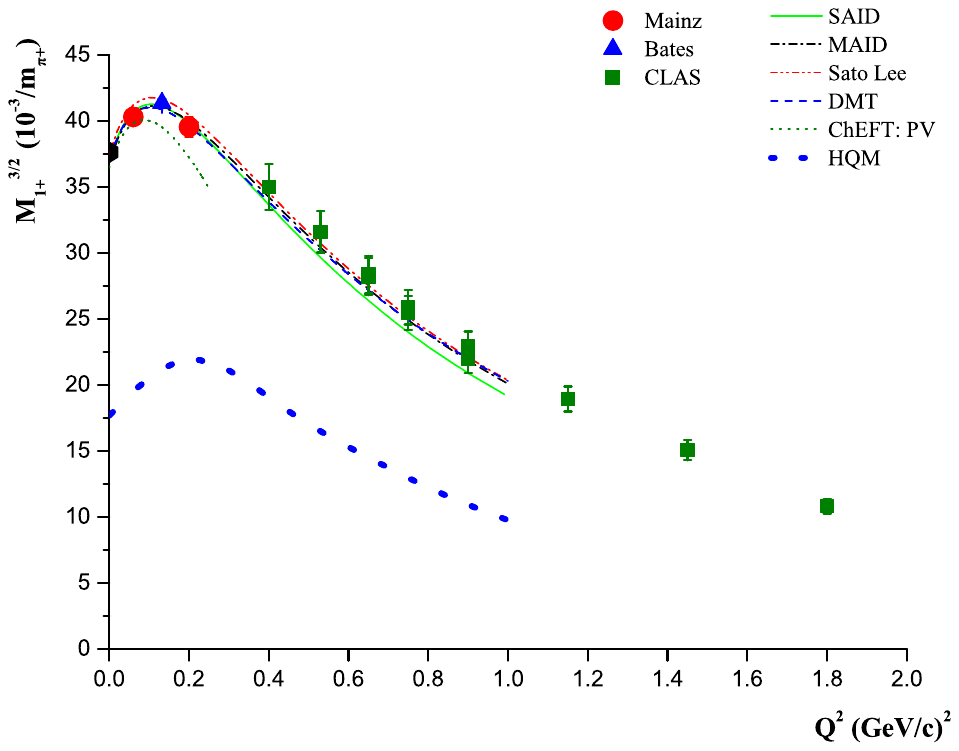}\\
\vspace*{-1cm}
\hspace*{-2cm}\includegraphics[width =8.5cm]{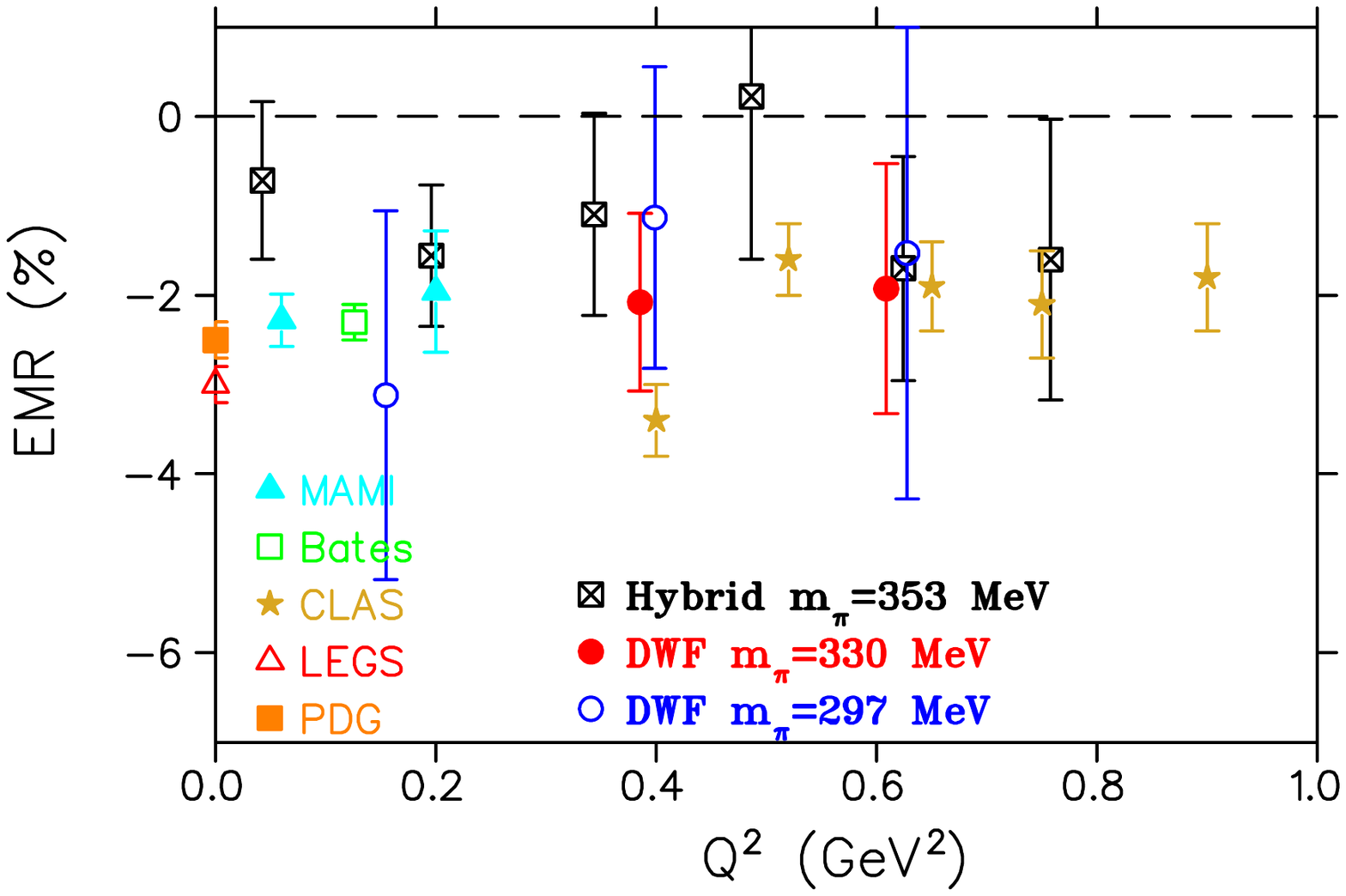}\\
\vspace*{-6cm}
\hspace*{-2cm}\includegraphics[width =8.5cm]{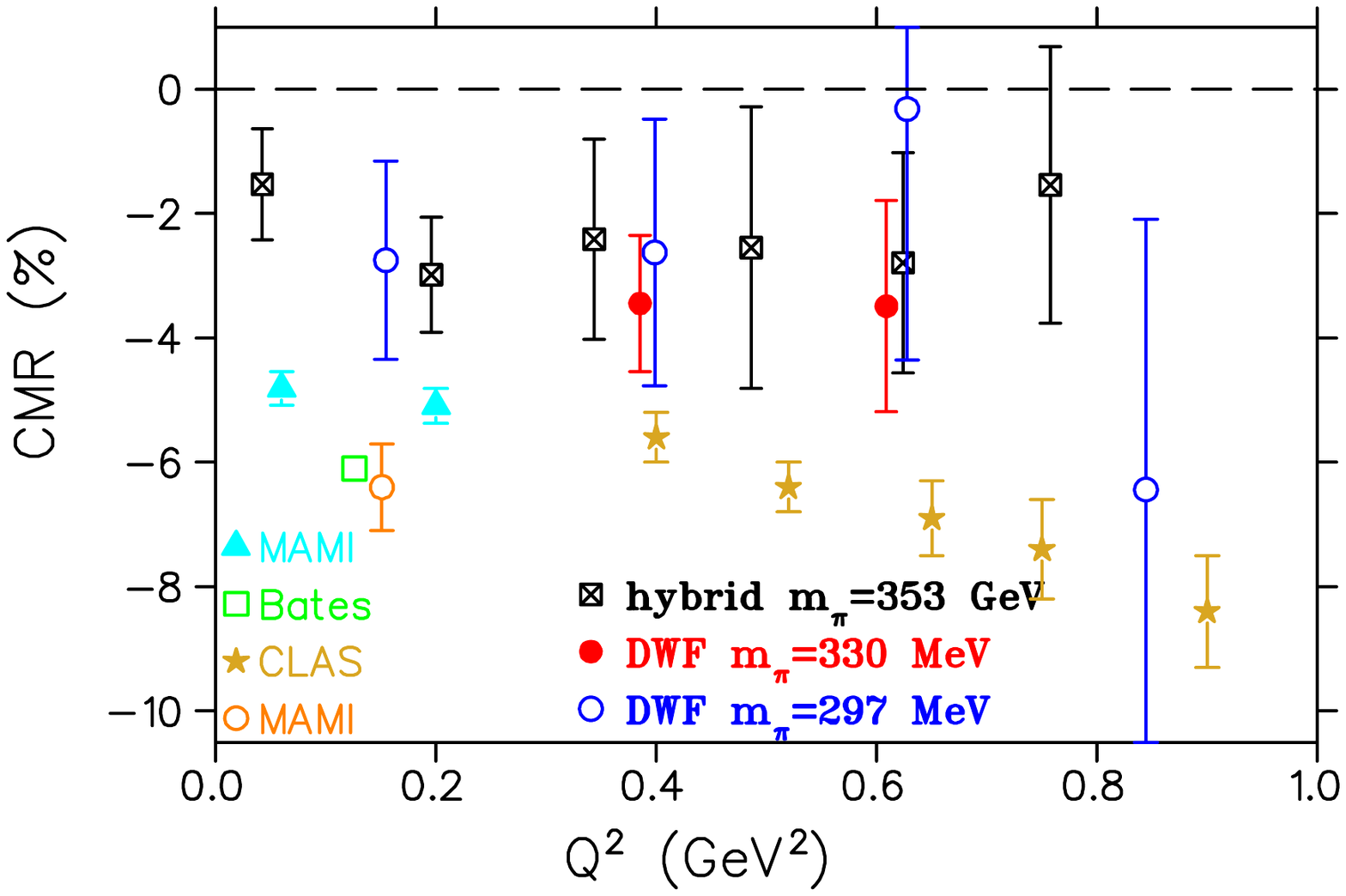}
\vspace{-5cm}
\caption{ 
The experimentally derived values for  $M^{3/2}_{1+}$ (M1) (top panel), CMR (middle panel)
and EMR (lower panel) compared to phenomenological  model results for M1, and to lattice for
EMR and CMR.
The derived multipole ratios are  shown
without the model error that is of the order or larger than the
depicted experimental error.} \label{fig:M1CMREMR}
\end{center}
\vspace*{-0.5cm}
\end{figure}

In the recent electroproduction experiments, which almost invariably
are carried out with polarized beams,
the transverse-longitudinal response functions $\sigma_{\rm TL}$ and $\sigma_{\rm TL'}$
are measured. Their simultaneous measurement 
 allows
the extraction of the real and imaginary parts of the same combination of multipole amplitudes.
Knowledge of both responses is particularly valuable because $\sigma_{\rm
TL}$ is most sensitive to the presence of a resonant longitudinal quadrupole
amplitude, while $\sigma_{\rm TL'}$ is particularly sensitive to the
background contributions, thus providing information on the two aspects of the problem that need to be controlled
independently~\cite{Mandeville:1994mx}. The importance of background is clearly seen in the W
behavior of the responses ~\cite{Mertz:1999hp} and the non-vanishing
recoil polarization $P_n$~\cite{Warren:1999pq, Pospischil:2000ad}, which bears 
close resemblance to $\sigma_{\rm TL'}$.
The transverse-transverse response,  $\sigma_{\rm TT}$, which is sensitive to the
electric quadrupole amplitude, was only recently
isolated for the first time at non-zero $Q^2$, with experiments pursued at Bates, JLab and
MAMI~\cite{Sparveris:2007zz,Cole:2007zz}.

Fig.~\ref{fig:M1CMREMR} offers a 
compilation  of CMR and EMR as a function of
$Q^2$. Both EMR and CMR  are small and
negative in the region where they have been measured.
From the accuracy of the present data one immediately
recognizes that the quark model predictions, which historically provided the motivation
for these measurements, do not agree with the data. In particular the dominant M1 matrix element
is found to be $\simeq$ 30\% stronger and the E2 and C2 amplitudes
at least an order of magnitude larger and often of
a different sign than the predictions of quark models.
This failure is to be expected since the quark model does not respect chiral symmetry
whose dynamic breaking leads to a strong, non-spherical, pion cloud
surrounding hadrons~\cite{amb}. As
will be discussed in detail in Section~\ref{sec:theory}, it was realized
that the pion cloud was a necessary ingredient to be added to quark models.
This is demonstrated by the calculations of the SL~\cite{SL} and DMT~\cite{DMT}  models,
which describe the data adequately, and which
show that most of the strength of the responses (and the EMR and CMR values) at very low $Q^2$
values, below $\simeq 0.25$~GeV$^2$/c$^2$, arises on account of the
mesonic degrees of freedom.
The recent results from MAMI along with the earlier ones from Bates
~\cite{Sparveris:2007zz} and the recent low $Q^2$ measurements from
CLAS~\cite{Cole:2007zz}, give strong support to this interpretation.
At asymptotic values of $Q^2$ helicity conservation~\cite{Carlson:1985mm} requires that EMR
$\rightarrow 1$ and that CMR $\rightarrow $~constant. Clearly this
regime has not been reached. The upgrade of CEBAF to 12 GeV will allow to extend
the measurements to higher $Q^2$, although
this will pose significant challenges in isolating the relevant
partial cross sections and even bigger ones in extracting the
relevant amplitudes.

Finally, the  $H (\vec{e},e'p)
~\gamma$ channel (VCS), which only recently has been accessed, allows the extraction
of the quadrupole amplitudes through a purely electromagnetic reaction channel providing an important cross
check to  the derived results from the pionic channel.
The  dispersion theory of Ref.~\cite{Pasquini:2001yy},
allows one to address the physics of deformation and
of nucleon polarizabilities in the region above pion threshold
simultaneously. Recent results from MAMI report the extraction
of polarizabilities~\cite{Bensafa:2006wr} and the first observation of
VCS data sensitive to the resonant quadrupole amplitudes~\cite{Sparveris:2008jx}. The results are in excellent agreement with those
derived from the pion channel.

\subsection{Sensitivity, Precision and Estimation of Uncertainties}

The  \GNdelta~data up to  $Q^{2}= 6.0$ \gevc~ are, in
general, characterized by small systematic errors and high statistical
precision. The interpretation of the data in terms of the  deformation
has been demonstrated, and as a result, the
research thrust shifted from the investigation of whether the
conjecture for deformation is valid to the exploration of the
mechanisms that cause it.  Investigating the physical origin
of deformation requires the measurement of new responses and the comparison of the theoretical results with
the experimentally derived quantities,  at a  level of precision far superior to the one feasible today.
This detailed comparison necessitates a reliable determination of the uncertainties of both the experimental results and the theoretical calculations.

The need for a critical and precise comparison of data and theory when extracting multipoles
in nucleon resonance studies is reminiscent of the ``crisis" in the
analysis of electron scattering data in the early 1970s,
where the very precise data could not be meaningfully compared with the theoretical calculations
in order to derive nuclear charge densities. This was primarily due to the lack of an appropriate
methodology that could enable to quantify the uncertainties
in the extracted densities, which, like multipole amplitudes, are not experimental observables.
The resolution of the ``crisis" through the introduction of a ``Model
Independent" extraction of charge densities led to a revolution in
the field and to the outstanding achievements in electron scattering. 

The leading method of extraction of multipole amplitudes, the Model
Dependent Extraction (MDE), produces extracted values that are
biased by the model and characterized by a  model
error, which is hard to estimate, especially if a single model is
employed~\cite{Frolov:1998pw,Pospischil:2000ad,Joo:2001tw,Elsner:2005cz,Ungaro:2006df}. An {\it Ansatz} for estimating the model uncertainties in the
extracted multipoles has been proposed~\cite{cnpepja,Stave:2007et} and used
in a few cases ~\cite{Sparveris:2004jn, Stave:2006ea}. In this method the same data
are analyzed employing different models, which describe the data
adequately, and attributing the resulting spread in the extracted
quantities to model uncertainty. Even though the phenomenological models available are of
considerable sophistication, 
  the small non-resonant amplitudes collectively could induce
large correlations and error in the extraction of the resonant amplitudes, resulting in the unsatisfactory  situation that the uncertainty is only approximately known.

A novel model independent method, the Athens Model Independent Analysis Scheme (AMIAS), for extracting multipole information from experimental nucleon
 resonance~\cite{Stiliaris:2007ru} and for analyzing lattice QCD simulation data has been presented ~\cite{Alexandrou:2008bp}. The method quantifies the uncertainty of the extracted multipoles and yields new information on background amplitudes, which MDE is incapable of accessing~\cite{Stave:2007et}.  Results from AMIAS are
shown in Fig.~\ref{fig:errorbands} (yellow bands) for the CMR sensitive \sLT~ partial cross
section with the precisely defined one-$\sigma$  uncertainty.
The experimentally allowed \sLT~  partial cross section,
as constrained by the Bates and MAMI data at $Q^{2}=0.127$ \gevc , are
shown as a function of $\theta^*_{pq}$. They are compared with
theoretical model predictions that account for them. It is evident
that the model predictions (dotted curves) with resonant quadrupole
amplitudes set to zero, which amounts to spherical solutions, are
excluded with high confidence.
On the contrary  model predictions from models that allow mesonic degrees of freedom allowing for deformation are in reasonable agreement with the experimental results at the $2\sigma$ level. Differences among the curves  predicted by the  various phenomenological models
are visible, but no inference can be drawn as their model error is not  known.
\emph{Nevertheless,  the comparison
demonstrates with extremely high confidence, with experimental errors precisely defined, that the assumption of
sphericity for both the nucleon and the $\Delta^{+}(1232)$ is
incompatible with the data.}


\section{The shape of nucleon and $\Delta$ resonance : theoretical understanding}
\label{sec:theory}

Having seen first experimental evidence for a non-spherical shape 
of the nucleon and $\Delta$ resonance, we next discuss its theoretical understanding. 
For the $\Delta$ resonance, its e.m. FFs are not accessible experimentally. They
are, therefore, an ideal example of observables where lattice QCD 
can make predictions.
 The state-of-the-art of these calculations as well as
 their implication on the shape of the $\Delta$ resonance are discussed.
 Subsequently our current theoretical understanding of the $\gamma^* N \to \Delta$
 transition is summarised
including an interpretation of the data presented in Section~\ref{sec:experimental}.

\subsection{$\Delta$ charge densities : lattice QCD}

As the nucleon is a spin-1/2 particle, its transverse charge densities
do not exhibit a quadrupole pattern, nor do they encode any information on its shape.
For spin-3/2 baryons, such information can however be obtained from the
 charge densities.

The matrix element of the e.m. current operator $J^\mu$
between spin-3/2 states, such as the $\Delta(1232)$ resonance,
can be decomposed into four multipole transitions:
Coulomb monopole (E0), magnetic dipole (M1),
Coulomb quadrupole (E2) and
magnetic octupole (M3), described by the
corresponding FFs $G_{E0}$, $G_{M1}$, $G_{E2}$ and $G_{M3}$~\cite{Nozawa:1990gt,Pascalutsa:2006up}.
Their values at $Q^2 = 0$ define e.g. the magnetic dipole moment~: 
$\mu_\Delta = G_{M1}(0)  e / (2 M_\Delta)$,
or the electric quadrupole moment~: 
$Q_\Delta = G_{E2}(0) e /  M_\Delta^2$.

The empirical knowledge of the $\Delta$ electromagnetic moments
is scarce, even though there were several attempts to measure its
magnetic dipole moment.
The current PDG value of the $\Delta^+$
magnetic dipole moment is~\cite{Nakamura:2010zzi}:
\beq
\mu_{\Delta^+} =  2.7 \mbox{${{+1.0} \atop {-1.3}}$}
(\mathrm{stat.}) \pm 1.5 (\mathrm{syst.}) \pm 3 (\mathrm{theor.})\, \mu_N\,.
\label{eq:mdmex}
\eeq
 This result was  obtained from
{\it radiative photoproduction}~($\gamma N \to \pi N \gamma^\prime$) of
neutral pions in the $\Delta(1232)$ region by the TAPS Collaboration at
MAMI~\cite{Kotulla:2002cg}, using
a phenomenological model of the $\gamma p \to \pi^0 p \gamma^\prime$
reaction.
For the $\Delta^+$, Eq.~(\ref{eq:mdmex}) implies~:
\begin{equation}
G_{M1}(0) =  3.5 \mbox{${{+1.3} \atop {-1.7}}$}
(\mathrm{stat.}) \pm 2.0 (\mathrm{syst.}) \pm 3.9 (\mathrm{theor.})\, .
\end{equation}
The size of the error-bar is rather large due to both experimental and theoretical uncertainties.

For the $\Delta$ electric quadrupole moment or magnetic octupole moments,
no direct measurements exist, nor do we have any empirical information on the $Q^2$ behavior
of the $\Delta$ e.m. FFs.
We thus rely on recent lattice QCD
calculations~\cite{Alexandrou:2009hs}
that can predict these FFs.

Calculation of the  e.m. FFs within lattice QCD requires   the evaluation of a three-point function,
as depicted in Fig.~\ref{fig:diagrams}.
We only consider here the connected diagram. Its evaluation
involves  two spatial sums :
one over the spatial coordinates of the operator and one over the
spatial coordinates of the final state.  In the so-called fixed sink method,
the sum over $\vec{x}_2$ is done automatically by generating a sequential (backward)
propagator  from the sink to the operator.
 Inserting the operator, which can be done at all values of $\vec{x}_1$,
and summing over $\vec{x}_1$ with the appropriate Fourier phase
  and propagator starting at $t=0$ and
ending at $t=t_1$ yields the connected three-point function,
for all momentum transfers $\vec{q}$.
To extract the matrix element  $\langle h^\prime (p^\prime)|{\cal O}|h(p) \rangle $  one
studies the large $t_1$ Euclidean time behavior of an appropriately
defined ratio of the three-point function and two-point functions, 
yielding a time independent quantity (plateau). Such a behavior signals
identification of the lowest hadron states $h$ and $h^\prime$ from
the tower of QCD states with the same quantum numbers
 as $h$ and $h^\prime$. Fitting to this plateau value we 
extract the matrix element $\langle h^\prime(p^\prime) |{\cal O}|h(p)\rangle $
and from this, depending on the choice of ${\cal O}$, the FFs or moments of parton distributions.

\begin{figure}[h]\vspace*{-1.5cm}
\begin{center}
\includegraphics[width =0.8\linewidth,height=.84\linewidth]{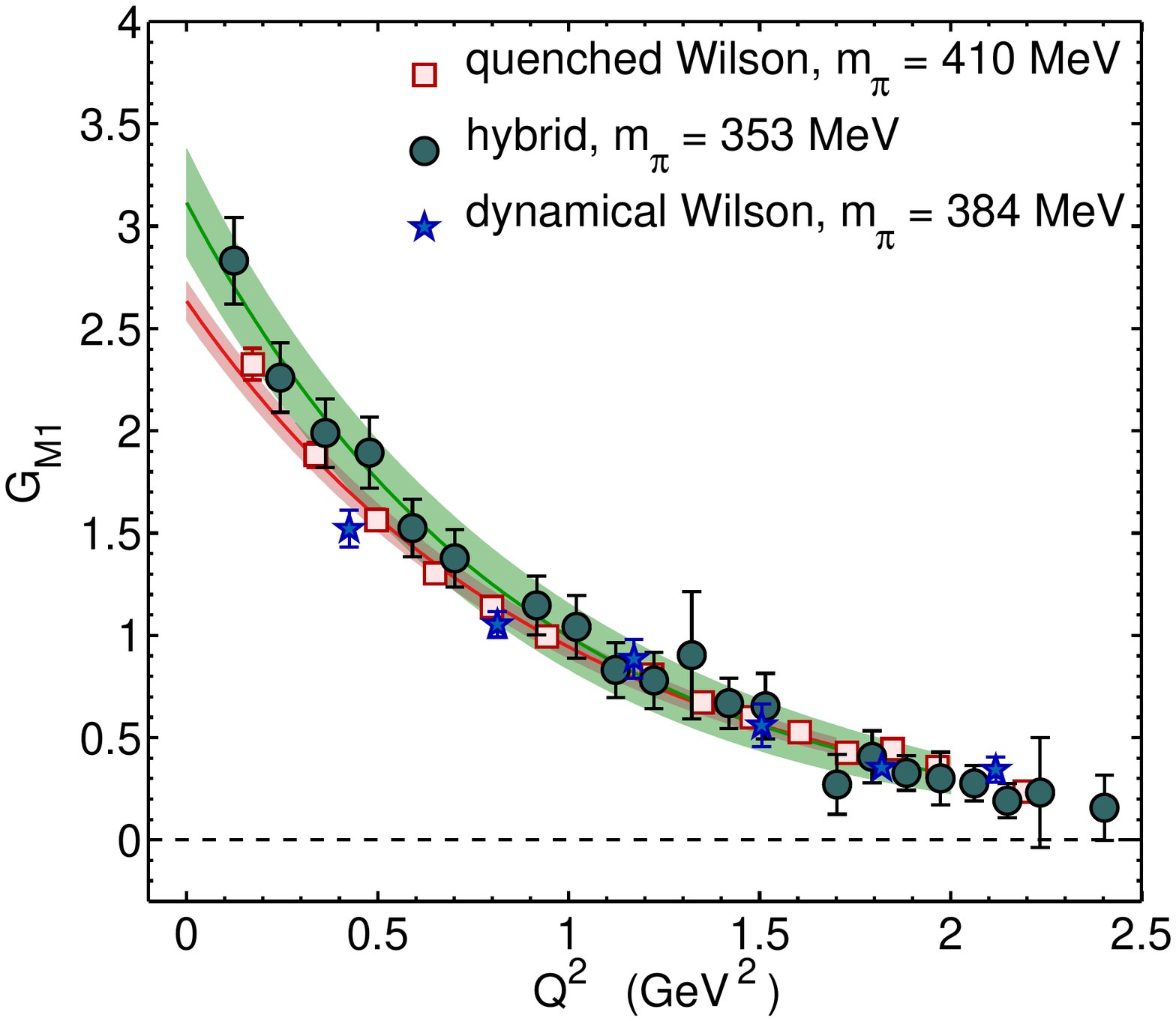}\\
\vspace*{-4cm}
\includegraphics[width =0.8\linewidth]{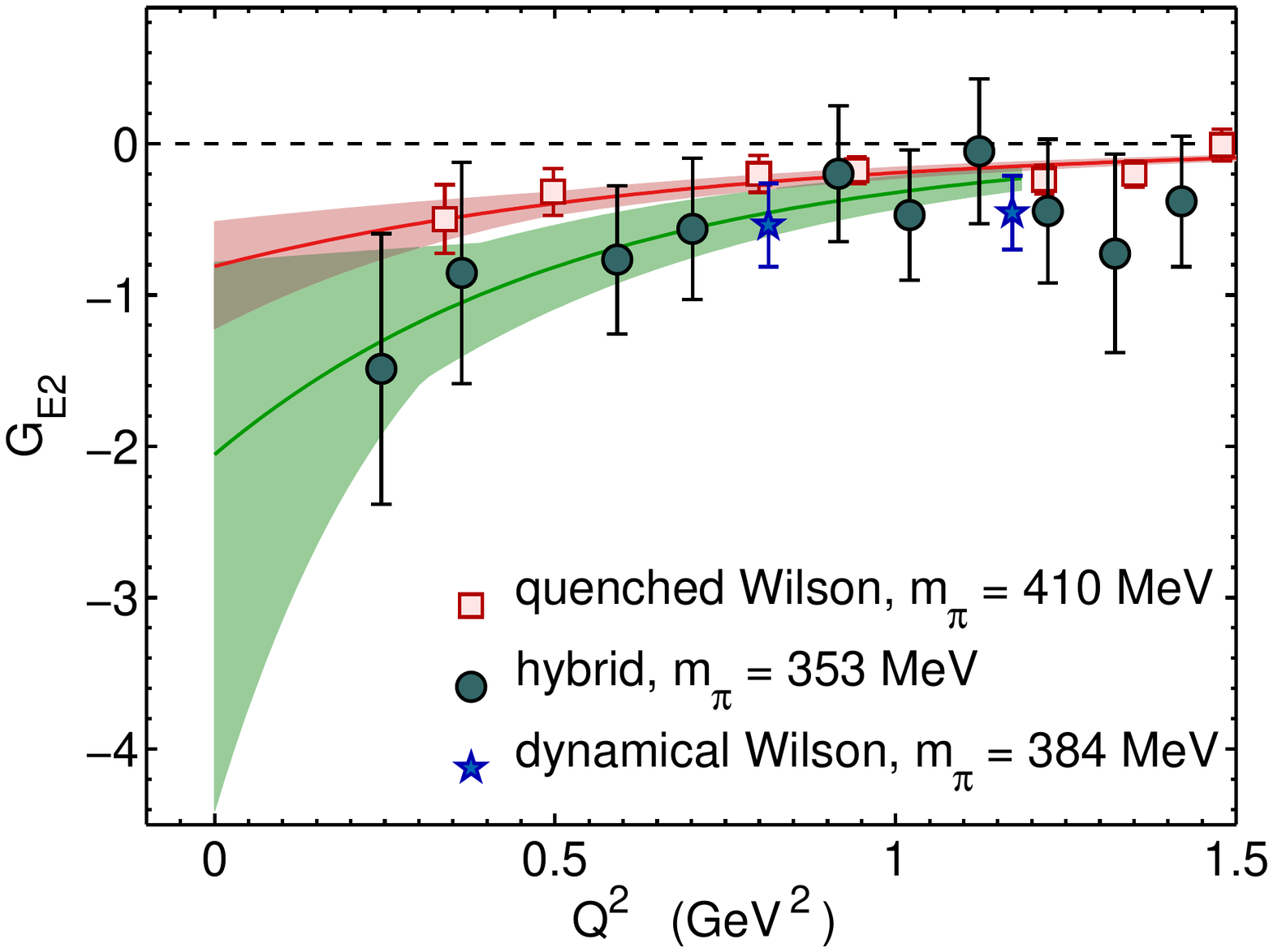}
\vspace{-2cm}
\caption{
Lattice results for the
$\Delta^+(1232)$  form factor
 $G_{M1}$ (upper panel) and $G_{E2}$ (lower panel) at the smallest pion mass
in three simulations~\cite{Alexandrou:2009hs}.
 The lines show the fits to an exponential form  of the quenched lattice results
and to the results obtained using the hybrid action.
The error band is calculated using a jackknife analysis on the fitted parameters.
}
\label{fig:gm1}
\end{center}
\vspace*{-0.5cm}
\end{figure}

To probe hadron deformation we will use in the following the
e.m. current as the operator ${\cal O}$.
Since the connected diagram for the $\Delta$ e.m. FFs is calculated by performing sequential inversions through
the sink,  the initial and final $\Delta$ states need to be fixed.
The $\Delta$ is described by a Rarita-Schwinger spinor and therefore there is some
freedom in the vector indices that can be chosen.
Ref.~\cite{Alexandrou:2009hs} concentrates on a few
carefully chosen combinations that best determine the three
FFs paying particular attention in constructing
a combination that isolates the electric quadrupole FF.
In order to efficiently check the lattice set-up, a quenched calculation is carried out
 using
Wilson fermions and the standard Wilson plaquette gauge action~\cite{Alexandrou:2009hs} for which statistical fluctuations are small.
Quenched results are then compared to a calculation using
 two dynamical degenerate flavors of Wilson fermions ($N_F=2$)
and the standard Wilson plaquette
gauge-action as well as using a hybrid action~\cite{Alexandrou:2009hs}.
The latter case uses two degenerate flavors of light
staggered sea quarks
and a  strange staggered sea quark ($N_F=2+1$) simulated using the Asqtad MILC
action~\cite{Bernard:2001MILC}.
The strange quark mass is
fixed to its physical value. These gauge configurations
are among the best simulations of the QCD vacuum  available. 
The valence quarks are 
domain wall fermions (DWF) that preserve a form of chiral symmetry on the
lattice.
A comparison between results obtained with these two different 
lattice formulations for the quarks (i.e. dynamical Wilson and staggered sea with DWF) provides
a  non-trivial check of lattice artifacts.  In both dynamical simulations  the
$\Delta$ is a stable particle.

We show the results for the $\Delta$ FFs $G_{M1}$ and $G_{E2}$ in Fig.~\ref{fig:gm1}.
For the pion masses considered,
there is agreement among results using the different actions,
with statistical errors being smallest in the quenched theory, as expected.
For the $\Delta$ magnetic dipole moment, first dynamical results,
using a background field method, with $N_F=2+1$ quark flavors
were presented in Ref.~\cite{Aubin:2008qp}. The magnetic moment can also be
extracted by fitting the $Q^2$-dependence of
 the magnetic dipole form factor $G_{M1}$ to determine its value at $Q^2=0$.
The values obtained in these two approaches are in agreement~\cite{Alexandrou:2009hs}.

Having a determination of the $\Delta$ e.m. FFs in lattice QCD
one can calculate its transverse light-front charge density
$\rho^\Delta_{T \, s_\perp}$~\cite{Alexandrou:2009hs},
as shown in Fig.~\ref{fig:deltatrans}.
Choosing the transverse spin vector $\vec S_\perp = \hat e_x$, the electric quadrupole
moment in a state of $s_\perp = +3/2$ for such charge distribution is then obtained from Eq.~(\ref{eq:dens3}) as~:
\be
Q_{+\frac{3}{2}}
= \frac{1}{2}  \left[ 2 \left( G_{M1}(0) - 3 e_\Delta \right)
+\left( G_{E2}(0) + 3 e_\Delta \right) \right]
 \left( \frac{e }{ M_\Delta^2} \right).
\label{eq:delquadrup}
\ee
Note that for a spin-3/2 particle without internal structure,
for which $G_{M1}(0) = 3 e_\Delta$ and $G_{E2}(0) = -3 e_\Delta$, the quadrupole moment of its
transverse light-front charge density vanishes.
This is in contrast with the non-relativistic case, where a non-zero value of $G_{E2}$ 
is usually interpreted as a non-zero quadrupole moment in the lab frame.
It is thus interesting to observe from Eq.~(\ref{eq:delquadrup})
that, as for the case of a spin-1 particle discussed in Section II,  
$Q_{s_\perp}$ is only sensitive to the anomalous parts
of the spin-3/2 magnetic dipole and electric quadrupole moments,
and vanishes for a particle without internal structure. 
Extrapolating the $\Delta$ e.m. FFs to $Q^2$ = 0, and using the extracted values in
Eq.~(\ref{eq:delquadrup}) yields a 
 quadrupole moment  $Q_{+\frac{3}{2}} $,
 of  $(0.73 \pm 0.16)$~$( e / M_\Delta^2)$ for the quenched 
  and $(0.51 \pm 0.22)$~$(e / M_\Delta^2 )$ 
for the hybrid cases. Both calculations therefore show a (small) prolate deformation of the two-dimensional
light-front charge density along the axis of the $\Delta$ spin (for the case of spin projection +3/2).

\subsection{The electromagnetic $N \to \Delta$ transition in QCD}

\subsubsection{Electromagnetic moments and densities}

Direct experimental evidence for a deformation of $N$ and $\Delta$
states can be obtained from the $\gamma^* N \Delta$ transition, which is usually characterized
in terms of three Jones--Scadron FFs~\cite{Jones:1972ky}~: $G^*_{M1}$,  $G^*_{E2}$ and
$G^*_{C2}$, denoting the magnetic dipole, electric quadrupole and Coulomb quadrupole
transitions respectively. For a review and more details see Ref.~\cite{Pascalutsa:2006up}.
In the following, we will often discuss the ratios EMR and CMR,  which
are expressed in terms of the Jones-Scadron FFs as:
\begin{eqnarray}
\mathrm{EMR} = - \frac{G_{E2}^\ast}{ G_{M1}^\ast} \,, \quad \quad
\mathrm{CMR} = - \frac{Q_+ Q_-}{4M_\Delta^2} \, \frac{G_{C2}^\ast}{G_{M1}^\ast},
\label{eq:ratiosjs}
\end{eqnarray}
with $Q_\pm \equiv \sqrt{(M_\Delta \pm M_N)^2 +Q^2}$.

\begin{figure}[h]\vspace*{-4cm}
\begin{center}
\includegraphics[width =7.25cm]{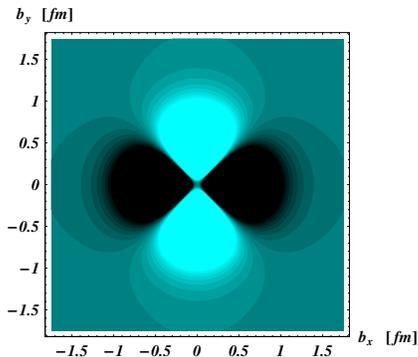}
\end{center}
\vspace{-0.75cm}
\caption{Quadrupole contribution to the transverse charge density
for the $p \to \Delta^+$ transition,
when $N$ and $\Delta$ are polarized along the $x$-axis with spin projection +1/2.
For the $N \to \Delta$ e.m. FFs, the phenomenological MAID2007~\cite{Drechsel:2007if} parametrization
is used. Figure from Ref.~\cite{Carlson:2007xd}.}
\label{fig:ndeltadens}
\vspace*{-0.5cm}
\end{figure}

From the experimental information on the $\gamma^* N \Delta$ transition,
discussed in Section~\ref{sec:experimental}, one can extract the transition magnetic dipole
and electric quadrupole moments from the values of the FFs at $Q^2 = 0$~\cite{Tiator:2003xr}:
\begin{eqnarray}
\mu_{N \to \Delta} &=& \sqrt{M_\Delta / M_N} \, G_{M1}^\ast(0) \quad [ \mu_N ] ,
\label{eq:mundel} \\
Q_{N \to \Delta} &=& - 6 \sqrt{M_\Delta / M_N}
\frac{2 M_\Delta}{M_N  (M_\Delta^2 - M_N^2)}  G_{E2}^\ast(0). \;
\label{eq:qndel}
\end{eqnarray}
Using the experimental information,  this yields~\cite{Tiator:2003xr}~:
\begin{eqnarray}
\mu_{p \to \Delta^+} \,&=&\, \left[ 3.46 \pm 0.03 \right ] \mu_N,
\label{eq:mundelexp} \\
Q_{p \to \Delta^+} \,&=&\, - \left( 0.0846 \pm 0.0033 \right )
\; \mathrm{fm}^2.
\label{eq:qndelexp}
\end{eqnarray}
\indent
One often uses an equivalent  parametrization for the $\gamma N \Delta$
transition at the real photon point ($Q^2 = 0$)
through two helicity amplitudes $A_{1/2}$ and $A_{3/2}$, where the subscript denotes the
total $\gamma + N$ helicity in the $\Delta$ rest frame.
Furthermore, one can generalize the considerations for the nucleon and $\Delta$
FFs to extract from the empirical information on the $Q^2$ dependence of the
$M1$, $E2$, and $C2$ transition FFs, the quark transition charge densities in the transverse plane, which induce the e.m. $N \to \Delta$ excitation~\cite{Carlson:2007xd}.
The transition density in a transversely polarized $N$ and $\Delta$
shows both monopole, dipole, and quadrupole patterns. The latter, shown
in Fig.~\ref{fig:ndeltadens}, maps the spatial dependence in the deformation of the transition charge distribution.

\subsubsection{Model descriptions of the $\gamma^* N \Delta$ transition }

As discussed above, the e.m. $N \to \Delta$
transition is predominantly of the magnetic dipole ($M1$) type.
A first understanding of the $\gamma^* N \Delta$ transition can be obtained
based on symmetries of QCD and its large number-of-color ($N_c$) limit.
In this limit, the baryon sector composed of up, down, and strange quark flavors of QCD
displays an $SU(6)$ spin-flavor symmetry.
This spin-flavor global symmetry of QCD is at the basis of many quark models, in which
baryons are described as (non-relativistic)
quantum-mechanical three-quark systems moving in a confining potential.
In such quark-model picture,
the $N\to \Delta$ transition is described by an $M1$ spin flip of a quark in the $S$-wave state,
illustrated in Fig.~\ref{fig:m1e2ndel}. 
The SU(6) symmetry allows to relate the magnetic dipole
moments of the proton and the $p \to \Delta^+$ transition as~:
$\mu_{p \to \Delta^+} = 2 \sqrt{2} / 3 \, \mu_p  =  2.63 \, \mu_N$,
which is about 25 \% lower than the experimental number of Eq.~(\ref{eq:mundelexp}).
Any $D$-wave admixture
in the nucleon {\it or} the $\Delta$ wave functions allows non-zero values
for the $E2$ and $C2$ quadrupole transitions, as illustrated in Fig.~\ref{fig:m1e2ndel}.

The prototype quark model is the Isgur-Karl model~\cite{Isgur:1978xj},
where the constituent quarks  move in a harmonic oscillator type long-range confining potential,
which is supplemented by an interquark force corresponding with one-gluon exchange.
The one-gluon exchange leads to a color hyperfine interaction,
which was found to predict well the mass splittings between octet
and decuplet baryons~\cite{DeRujula:1975ge}.
This  hyperfine interaction contains a tensor force
which produces a $D$-state admixture in the
$N$ and $\Delta$ ground states, around
1~\% ~\cite{Koniuk:1979vy,Isgur:1981yz}.
As a result of such $D$-wave components, the $N$ and $\Delta$
charge densities become non-spherical, yielding small
negative EMR values, in the range
 $-0.8 \, \% <$~EMR~$< -0.3 \, \%$ within non-relativistic quark models~\cite{Isgur:1981yz, Gershtein:1981zf}.
The small value for EMR already indicates that any effect of deformation in
the nucleon and/or $\Delta$ ground state is rather small and very sensitive
to details of the wave function, as well as 
truncation in the quark model basis~\cite{Drechsel:1984ie,Giannini:1990pc}.
The error induced due to the truncation in the quark model basis has
been further investigated in the relativized quark model~\cite{Capstick:1989ck}, 
typically resulting in an even smaller negative value, namely EMR~$ \simeq -0.2 \%$.

Even though the constituent quark model, despite its simplicity,
is relatively successful in predicting the structure and spectrum of
low-lying baryons, it under-predicts $\mu_{N \to \Delta}$ by more than
25 \% and leads to values for  EMR, which are typically
smaller than experiment.
More generally, constituent quark models do not satisfy the symmetry
properties of the QCD Lagrangian. In the limit of massless up and down (current) quarks, 
the QCD Lagrangian is
invariant under $SU(2)_L \times SU(2)_R$ rotations of left ($L$) and
right ($R$) handed quarks in flavor space. This {\it chiral symmetry}
is spontaneously broken in nature leading to the appearance of massless
Goldstone modes, pions, which acquire a mass due to the explicit breaking of chiral
symmetry. 
Since pions are the lightest hadrons, they
dominate the long-distance behavior of hadron wave functions. 
As the $\Delta$(1232) resonance nearly entirely decays into $\pi N$, the
pions are of particular relevance
to the $\gamma^* N  \Delta$ transition.
Therefore, a natural way to qualitatively improve
on the above-mentioned constituent
quark models is to include the pionic degrees of freedom.

Early investigations of the $\gamma^* N \Delta$ transition including
pionic effects were performed within the
{\it chiral bag model}~\cite{Kaelbermann:1983zb, Bermuth:1988ms},
which was developed as an improvement to the MIT bag model~\cite{Donoghue:1975yg}
by introducing an elementary pion, which couples to quarks in the bag
in such a way that chiral symmetry is restored~\cite{Thomas:1982kv}.
Calculations within the chiral bag model~\cite{Lu:1996rj},
found that with a bag radius, $R$, around $0.8$ fm
one is able to obtain a reasonably good description for the
helicity amplitudes, as can be seen from the values given in  Table~\ref{table_emr}.
For such a small bag radius, the pionic effects
are crucial as they account for around 75~\% of the total strength of the
amplitude $A_{3/2}$. The same calculation however yields  EMR~$\simeq -0.03 \%$, in disagreement with  experiment.

The role of the pion-cloud contributions is also highlighted in
 {\it Skyrme models}~\cite{Wirzba:1986sc,Walliser:1996ps},
in which the nucleon appears
as a soliton solution of an effective non-linear meson field theory.
The inclusion of rotational corrections in such models,
leads to a quadrupole distortion of the classical soliton solution,
yielding a value for EMR~$ = -2.3 \%$~\cite{Walliser:1996ps}, consistent with experiment.

The EMR ratio has also been calculated in models, with both quarks and pion degrees of freedom such as the {\it chiral quark soliton model} ($\chi$QSM),
which interpolates between a constituent quark model and the
Skyrme model~\cite{Watabe:1995xy}.
For the two flavor case, one finds EMR~$ = -2.1 \%$~\cite{Silva:1999nz},
fairly close to experiment, considering that
in the $\chi$QSM calculation no parametrization adjustment has been made
to the $N \to \Delta$ transition.
However, the magnitudes of the photocouplings, which are given in Table~\ref{table_emr},
are largely under-predicted in the $\chi$QSM.

A number of subsequent works have revisited
quark models, restoring chiral symmetry by including 
two-body exchange currents between the quarks. These exchange currents
lead to non-vanishing $\gamma^* N \Delta$ quadrupole
amplitudes~\cite{Buchmann:1996bd},
even if the quark wave functions have no $D$-state
admixture. Such a picture~\cite{Buchmann:1996bd}, in which the $\Delta$
is excited by flipping the spins of two quarks, yields  
EMR~$ \simeq -3.5 \%$,  
and relates the $N \to \Delta$ and $\Delta^+$ quadrupole moments to the neutron
charge radius as~:
\begin{eqnarray}
Q_{p \to \Delta^+} =  r_n^2  / \sqrt{2}, \quad \quad \quad \quad
Q_{\Delta^+} = r_n^2 \, .
\label{eq:buchrel}
\end{eqnarray}
Using the experimental neutron charge radius, $r_n^2 = -0.113 (3)$~fm$^2$,
Eq.~(\ref{eq:buchrel}) yields~:
$Q_{p \to \Delta^+} = - 0.08 \; \mathrm{fm}^2$, and 
$Q_{\Delta^+} = - 0.113 \; \mathrm{fm}^2$. 
This value of $Q_{p \to \Delta^+}$ is close to the empirical determination, given in
Eq.~(\ref{eq:qndelexp}).
In such hybrid (quark/pion-cloud) models~\cite{Faessler:2006ky},
the pion cloud is fully responsible for the non-zero values of the intrinsic quadrupole moments
and hence for the non-spherical shape of these particles.
As a summary, we list in Table~\ref{table_emr}
the $\gamma N \Delta$ photo-couplings $A_{1/2}$ and $A_{3/2}$ as well as the
ratio EMR  in various models.

\begin{table}[h]
{\centering \begin{tabular}{|c|c|c|c|}
\hline
&
$A_{1/2}$  & $A_{3/2}$  & EMR  \\
&
$[10^{-3}\quad \quad $ &
$[10^{-3}\quad \quad $ & [\%]  \\
&
$~\mathrm{GeV}^{-1/2}]$ &
$\mathrm{GeV}^{-1/2}]$ &   \\
\hline
\hline
experiment &&& \\
Ref.~\cite{Nakamura:2010zzi} & $-135 \pm 6$ & $-250 \pm 8$ & $-2.5 \pm 0.5$  \\
\hline
\hline
SU(6) symmetry & -107 & -185 & 0 \\
\hline
quark models  &&& \\
non-rel..~\cite{Koniuk:1979vy,Isgur:1981yz,Gershtein:1981zf,Drechsel:1984ie}
& -103 & -179 & $-2 $  to  0 \\
relativized~\cite{Capstick:1989ck}
& -108 & -186 & $-0.2$ \\
\hline
bag models &&& \\
MIT~\cite{Donoghue:1975yg}
& -102 & -176 & 0 \\
chiral bag~\cite{Bermuth:1988ms}
& -106 & -198 & $-1.8$ \\
chiral bag~\cite{Lu:1996rj}
& -134 & -233 & $-0.03$ \\
\hline
Skyrme models \cite{Wirzba:1986sc} &&& \\
\cite{Walliser:1996ps}  & -136 & -259 & $-2.3$ \\
\hline
chiral quarks
& & &  \\
soliton~\cite{Silva:1999nz}   & -70.5 & -133 & -2.1 \\
$\pi, \sigma$ exchange~\cite{Buchmann:1996bd} & -91 & -182 & -3.5 \\
\cite{Faessler:2006ky} & -124.3 & -244.7 & -3.1 \\
\hline
\hline
\end{tabular}\par}
\caption{Summary of the $\gamma N \Delta$ photo-couplings
$A_{1/2}$, $A_{3/2}$, and EMR
in different models compared with experiment.}
\label{table_emr}
\end{table}

\subsubsection{Large $N_c$ predictions}

Although the results obtained from the different QCD inspired models reviewed above may
provide us with physical insight on the $\gamma^* N \Delta$ transition and its relation to the
nucleon and $\Delta$ shape, they are not a rigorous consequence of QCD.
In the following subsections, we
will discuss what is known on the $\gamma^* N \Delta$ transition
from approaches, which are directly related with  QCD in some limit,
such as the $1/N_c$ expansion of QCD (limit of large number of colors),
chiral effective field theory (chiral limit of small pion masses or momentum transfers) or
lattice QCD simulations (continuum limit).

The $1/N_c$ expansion of QCD~\cite{'tHooft:1973jz, Witten:1979kh}
provides an expansion with a perturbative parameter at all energy scales.
This expansion has proved quite useful in describing properties
of baryons, such as, ground-state and excited  masses, magnetic moments, and
electromagnetic decays.
For reviews see Refs.~\cite{Jenkins:1998wy,Lebed:1998st}.
For example, the $N \to \Delta$ transition
magnetic moment $\mu_{N \to \Delta}$ is related to the isovector nucleon magnetic moment as~\cite{Jenkins:1994md}:
$\mu_{p \to \Delta^+} = \left( \mu_p - \mu_n \right) / \sqrt{2} \simeq 3.23 \, \mu_N$, 
within 10 \% of the experimental value of Eq.~(\ref{eq:mundelexp}).
The EMR value was shown to
be of order $1/N_c^2$~\cite{Jenkins:2002rj}. Thus
its smallness  is naturally explained in the large $N_c$ limit.

The large $N_c$ limit also  allows one to relate the
$\Delta$ and $N \to \Delta$ quadrupole moments via~\cite{Buchmann:2002mm}~:
$Q_{\Delta^+} / Q_{p \to \Delta^+}  \,=\, 2 \sqrt{2} / 5 \,+\,
{\mathcal O} \left( 1 / N_c^2  \right)$,
Using the phenomenological value of Eq.~(\ref{eq:qndelexp})
yields~:
$Q_{\Delta^+} = - \left( 0.048 \pm 0.002 \right ) \; \mathrm{fm}^2$, 
which implies $G_{E2}(0) = -1.87 \pm 0.08$.

The relation of Eq.~(\ref{eq:buchrel}) between 
$Q_{p \to \Delta^+}$ and  $r_n^2$ was also shown~\cite{Buchmann:2002mm} 
to hold in the large $N_c$ limit. 
Furthermore, it was shown~\cite{Pascalutsa:2007wz} that
in the large $N_c$ limit~:
\begin{eqnarray}
\mathrm{EMR} = \mathrm{CMR} = (1/12) \,R_{N \Delta}^{3/2}\,(M_\Delta^2 - M_N^2) \, 
r_n^2 / \kappa_V,
\label{eq:rlnc}
\end{eqnarray}
with $R_{N \Delta} \equiv M_N/M_\Delta$, and $\kappa_V = \kappa_p - \kappa_n$,
the isovector nucleon anomalous magnetic moment.
Numerically, Eq.~(\ref{eq:rlnc}) yields
$\mathrm{EMR} = \mathrm{CMR} = - 2.77 \, \%$.
For EMR this
prediction is in an excellent agreement with experiment, Eq.~(\ref{eq:qndelexp}).
For CMR, where a direct measurement at the real-photon
point is not possible, extending the large-$N_c$ relation
to finite $Q^2$ allows relations with the
neutron electric FF, which agree well with experiment~\cite{Grabmayr:2001up, Pascalutsa:2007wz}.

\subsubsection{ Lattice QCD and chiral effective field theory}

\begin{figure}[h]\vspace*{-1cm}
\centerline{\includegraphics[width=\linewidth]{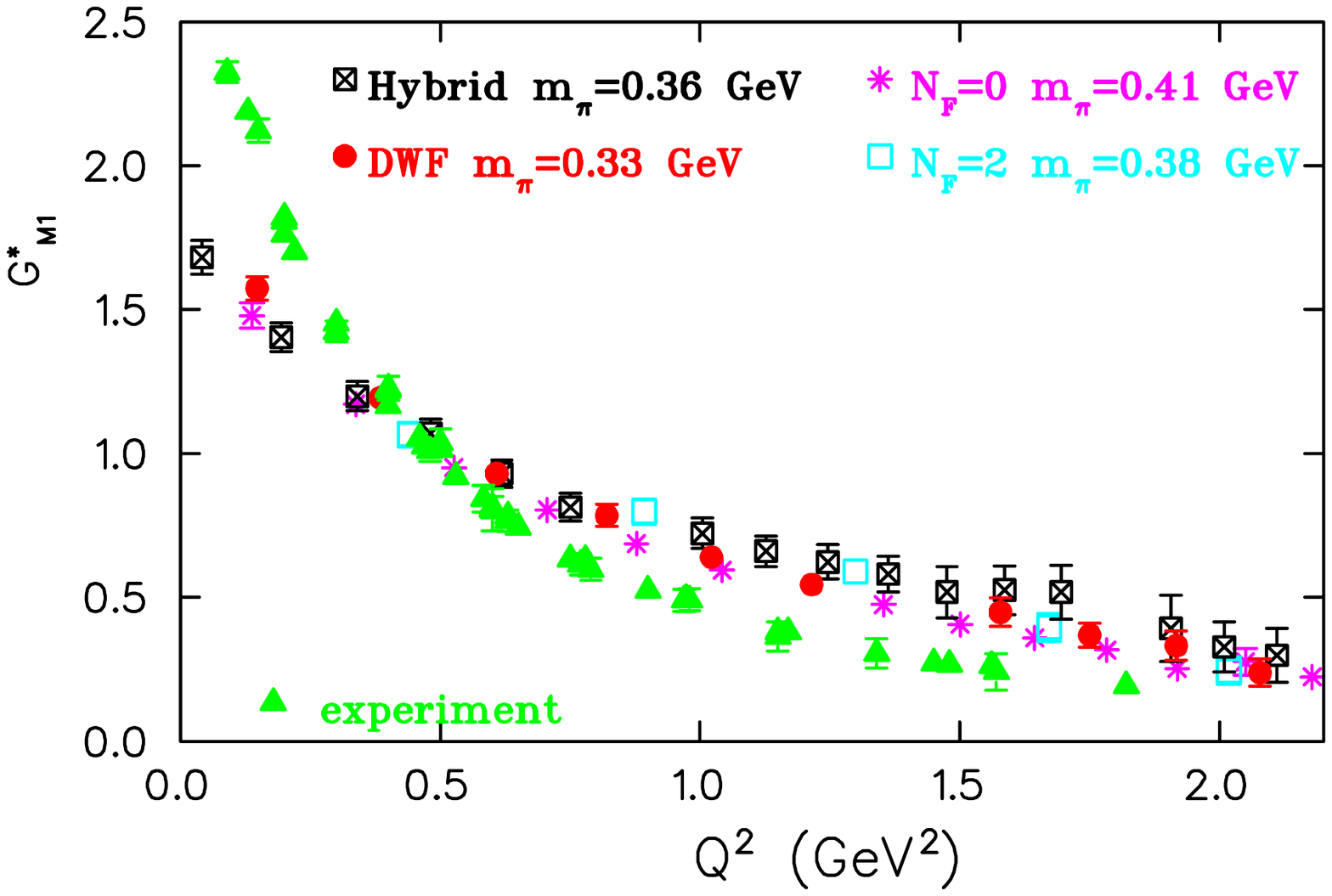}}
\vspace*{-5.5cm}
\caption{
$Q^2$-dependence of the $N \to \Delta$ FF $G^\star_{M1}$,
at the lightest pion mass for each type of simulation. Quenched results are  shown with
the asterisks, results with $N_F=2$ Wilson with the open squares, results using the hybrid action with the dotted squares and using DWF with the filled circles. Experimental data
from Refs.~\cite{Mertz:1999hp,Joo:2001tw,Beck:1997ew,Stave:2006ea,Sparveris:2006uk,GM1_exp} are shown with the filled triangles. Lattice data are from Refs.~\cite{Alexandrou:2007dt,Alexandrou:2010uk}.
}
\label{fig:latticeM1NDelta}
\end{figure}

Lattice QCD provides the possibility of calculating the $N$ to $\Delta$
e.m. FFs starting from the  underlying theory of QCD.
The set-up for the lattice calculation of the three  transition  FFs
is the same as that used for the extraction of the $\Delta$ FFs.
The advantage in this case is that
the connected diagram yields the full contribution. It is
calculated by sequential inversion through the sink for  the
same three simulations as described in the case of the $\Delta$ FFs.
In addition, recent calculations using $N_F = 2+1$ dynamical DWF
 generated by the  RBC and UKQCD collaborations~\cite{RBC}
 provide a unitary setup and a further check of the results~\cite{Alexandrou:2010uk}.
In all these calculations the pion mass is such that the $\Delta$ is still stable.
Fig.~\ref{fig:latticeM1NDelta} shows a comparison of the lattice
results for $G^\star_{M1}$ at the lightest pion mass in each type of simulation.
 There is agreement among  Wilson fermions and results obtained
 using the hybrid action  as well as dynamical DWF.
The agreement of results using dynamical fermions with the quenched results
indicate that pion cloud contributions due to pair creation are still small at
a pion mass of about 330~MeV.
As also seen for the nucleon e.m. FFs,   lattice results
 underestimate $G^\ast_{M1}$. Chiral dynamics is expected
to induce large corrections at small $Q^2$ and such effects can
be investigated as  lattice simulations
at smaller pion masses become available.

The CMR and EMR  are shown in Fig.~\ref{fig:M1CMREMR},
and have larger statistical errors due to the fact that $G^\star_{E2}$ and
$G^\star_{C2}$, being sub-dominant, are harder to determine. We show quenched results
that have the smallest errors as well as results obtained
in the hybrid action approach and using $N_F=2+1$ DWF~\cite{Alexandrou:2010uk}. The conclusion that can be drawn is that
agreement of lattice results on EMR and CMR with experiment
is better as compared to that of $G^\star_{M1}$. Such an agreement
is seen also in
other ratios, indicating that
they are less affected by lattice
artifacts than each of the quantities separately.

\begin{figure}[t]\vspace*{-1cm}
\includegraphics[width=.95\linewidth]{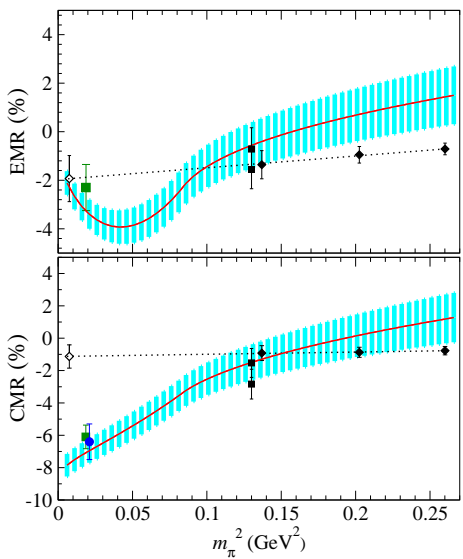}
\vspace*{-2.5cm}
\caption{
The $m_\pi$ dependence of
EMR (upper panel) and
CMR (lower panel), at $Q^2=0.1$ GeV$^2$.
The data points are from MAMI~\cite{Pospischil:2000ad}~(blue circle),
and BATES~\cite{Mertz:1999hp,Sparveris:2004jn})~(green squares).
The filled black diamonds are quenched lattice results~\cite{Alexandrou:2004xn}.
The open diamond near $m_\pi \simeq 0$
represents their extrapolation assuming linear dependence in $m_\pi^2$.
The solid black squares at $m_\pi = 0.36$~GeV are hybrid lattice
results~\cite{Alexandrou:2007dt} at $Q^2 \simeq 0.04$~GeV$^2$ (upper points) and
at $Q^2 \simeq 0.2$~GeV$^2$ (lower points).
The solid red curves are a $\chi$EFT result~\cite{Pascalutsa:2005ts}.
The (blue) error bands represent the estimate of theoretical uncertainty for
the $ \chi$EFT calculation.}
\label{fig:latticeextrap}
\end{figure}

The present lattice  QCD calculations are performed for quark masses
larger than their values in nature.
To extrapolate to the physical pion mass, one can use the
$\chi$EFT of QCD~\cite{Weinberg:1978kz,Gasser:1983yg,BKM}.
$\chi$EFT provides a firm theoretical framework at low scales, with
the relevant symmetries of QCD built in consistently.
The $\gamma^* N \Delta$ transition provides new challenges for
$\chi$EFT as it involves the interplay of
two light mass scales~: the pion mass and the $N - \Delta$ mass difference.
A first study, taking into account these two mass scales, was performed
within the framework of heavy-baryon chiral perturbation theory~\cite{Butler:1993ht}.
A more comprehensive study was subsequently carried out~\cite{Gellas:1998wx,Gail:2005gz} using
the ``$\epsilon$-expansion'' scheme.
In that scheme, the two light scales in the problem:
the pion mass $\epsilon \equiv m_\pi / \Lambda_{\chi \mathrm{SB}}$,
with $\Lambda_{\chi \mathrm{SB}} \sim 1$~GeV
the chiral symmetry breaking scale,
and the $\Delta$-resonance excitation energy
$\delta \equiv (M_\Delta - M_N) / \Lambda_{\chi \mathrm{SB}}$
are counted as being of the same order, i.e. $\epsilon \sim \delta$. 
To allow for an energy-dependent power-counting scheme designed to take 
account of the large variation of the $\Delta$-resonance contributions
with energy, the  ``$\delta$-expansion'' scheme has 
been introduced~\cite{Pascalutsa:2002pi}. 
It treats the two light scales $\epsilon$ and $\delta$
on a different footing, counting $\epsilon \sim \delta^2$,
the closest integer-power relation
between these parameters in the real world. 
It has been applied to the study of the  $\gamma^\ast N\Delta$ FFs~\cite{Pascalutsa:2005ts},  
and has been used in extrapolating
the present lattice QCD calculations to the physical pion mass.  This is shown in
Fig.~\ref{fig:latticeextrap} for the EMR and CMR ratios, which
shows that $\chi$EFT predicts strong non-analytic dependencies on the quark mass for $m_\pi <  (M_\Delta - M_N)$, invalidating simple linear extrapolation in $m_\pi^2$.
In particular, the $\chi$EFT results reconcile the lattice results and the relatively large negative experimental value for CMR.

For smaller pion masses, where the $\Delta$ becomes an 
open channel, the lattice results will be able to provide momentum-dependent 
phase-shifts~\cite{Luscher:1991cf}. To extract resonance quantities from those will require a fitting procedure, e.g. Breit-Wigner or 
complex pole fits, as done when extracting them from experimental multipoles.


\section{Conclusions}

In this review, we have presented the experimental results and theoretical
 understanding on the shape of hadrons. Although shapes of nuclei have been explored over many
 decades, it is only in recent years that
it became possible to define this question in a theoretically rigorous way for hadrons, and perform
the experiments to answer it.
The key concept is to quantify size and shape
of an extended object  through a quantum mechanical density operator.
For a relativistic bound state system of near massless quarks, a probability interpretation
is obtained by considering the system in a light-front frame, and projecting its charge density
along the line-of-sight. We have argued that the resulting
transverse charge density encodes the information on hadron size and shape.

 On the experimental side,  the most accessible and best studied  reaction to reveal
 hadron deformation is the $N \to \Delta$ transition.
We have reviewed the state-of-the-art experimental techniques, which have allowed to accurately determine the $N \to \Delta$ quadrupole amplitudes
 at low momentum
transfers, and establish a deformation in the $N / \Delta$ system. The quadrupole transitions were
pinned down on the order of a few \% of the dominant magnetic dipole transition.
A quantitative understanding of the small, non-zero values of these amplitudes from the underlying theory, QCD, is a particular challenge. We have provided the historical perspective in which
this question was addressed from QCD inspired models, highlighting the role the pions
play in these transitions. It is only very recently, however, that {\it ab initio} calculations became possible, and state-of-the-art full lattice QCD simulations for both the $N \to \Delta$ and $\Delta$ quadrupole FFs were able to quantify them. There is an ongoing effort by many groups to perform such simulations at pion masses approaching the physical value and reducing further lattice artifacts.

The theoretical foundations, experimental techniques, and lattice QCD simulation methods to access hadron deformation through the measurement of quadrupole FFs are well established now.
We can therefore expect in the near future that  a refinement of the lattice calculations as well as new high-precision experiments with polarized beams and polarized targets, will allow to further sharpen our understanding of hadron shapes.

\section*{Acknowledgements}

The work of C.A. is partly supported
by the Cyprus Research Promotion Foundation, of C.N.P. by the Ministry of Education, Greece and the Cyprus Institute, and of M.V. by the Research Center ``Elementary Forces \& Mathematical Foundations" at Mainz University.

We like to thank
A. Bernstein,
C. Carlson,
D. Drechsel,
P. Hoyer,
Th. Korzec,
G. Koutsou,
C. Lorc\'e,
V. Pascalutsa,
S. Stiliaris
N. Sparveris,
L. Tiator,
and
A. Tsapalis,
for useful discussions and correspondence.


\end{document}